\documentclass[%
 reprint,
 superscriptaddress,
 amsmath,amssymb,
 aps,
 prx,
]{revtex4-2}

\usepackage[colorlinks=true,
            linkcolor=blue,
            citecolor=blue,
            urlcolor=blue]{hyperref}

\usepackage{graphicx}
\usepackage{dcolumn}
\usepackage{bm}

\begin{document}

\preprint{APS/123-QED}

\title{Predicting the Interfacial Energy and Morphology of DNA Condensates}

\author{Sihan Liu}
 \affiliation{Department of Mechanical and Aerospace Engineering, Princeton University, Princeton, NJ 08544, USA}

\author{Andrej Ko\v{s}mrlj}
 \email{andrej@princeton.edu}
 \affiliation{Department of Mechanical and Aerospace Engineering, Princeton University, Princeton, NJ 08544, USA}
 \affiliation{Princeton Materials Institute, Princeton University, Princeton, NJ 08544, USA}

\date{\today}

\begin{abstract}
The physics and morphology of biomolecular condensates formed through liquid–liquid phase separation underpin diverse biological processes, exemplified by the nested organization of nucleoli that facilitates ribosome biogenesis. Here, we develop a theoretical and computational framework to understand and predict multiphase morphologies in DNA-nanostar solutions. Because morphology is governed by interfacial energies between coexisting phases, we combine Flory–Huggins theory with coarse-grained molecular dynamics simulations to examine how these energies depend on key microscopic features of DNA nanostars, including size, valence, bending rigidity, Debye screening length, binding strength, and sticky-end distribution. The phase behavior of DNA nanostars is quantitatively captured by a generalized lattice model, in which the interplay between sticky-end binding energy and conformational entropy determines the effective interactions. Focusing on condensates comprising two dense phases, we find that Janus-like morphologies are ubiquitous because the interfacial energies between the dense and dilute phases, \(\gamma_{i\in\{1,2\}}\), are typically comparable. In contrast, nested morphologies are rare as they require a large asymmetry in \(\gamma_i\), which arises only for highly dissimilar nanostars such as those differing markedly in valence or size. Moreover, the interfacial energy between the two dense phases, \(\gamma_{12}\), can be modulated either discretely, by varying sticky-end distribution, or continuously, by tuning the crosslinker ratio; the former may even eliminate nested configurations. These findings establish physical design principles for constructing complex condensate architectures directly from microscopic molecular parameters.
\end{abstract}

\maketitle

\section{Introduction}

The coexistence of multiple liquid phases is a ubiquitous physical phenomenon in both nature and industry. Examples include oil-in-water emulsions, beverages, medicines, and cosmetic products, all formed through liquid–liquid phase separation \cite{lohse2020physicochemical}. In biology, cells are inherently multicomponent liquid systems in which thousands of distinct biomolecules interact to perform diverse functions simultaneously—for instance, DNA transcription and replication, RNA processing, and protein synthesis \cite{brangwynne2015polymer, laflamme2020biomolecular, wang2021liquid, mittag2022conceptual}. Numerous biomolecular condensates within cells form through phase separation \cite{hyman2014liquid, boeynaems2018protein}, including nucleoli \cite{brangwynne2011active}, stress granules \cite{riback2017stress}, and P bodies \cite{parker2007p}. Enriched in polymers such as proteins and RNA, these liquid-like organelles compartmentalize the cell without membranes, thereby enabling the spatiotemporal regulation of internal components and diverse processes including gene transcription, RNA metabolism, and stress responses \cite{bracha2019probing, choi2020physical}.

Many biomolecular condensates contain multiple coexisting subphases separated by distinct interfaces \cite{gall1999assembly, shav2005dynamic, kedersha2005stress, feric2016coexisting, kaur2021sequence, lafontaine2021nucleolus, li2024predicting, grandpre2025membrane, yu2025pattern, quinodoz2025mapping}. These multiphase condensates exhibit a variety of morphologies \cite{boisvert2007multifunctional, decker2012p, jain2016atpase, feric2016coexisting, shin2017liquid, youn2019properties, sanders2020competing, lafontaine2021nucleolus}, governed by the interplay between interfacial energies among different phase pairs and the relative volume fractions of each phase \cite{mao2019phase, mao2020designing}. The morphology of such condensates provides an additional mechanism for spatially regulating biological processes, extending beyond the mere compartmentalization of essential components for specific biochemical reactions. A prime example is the “Russian-doll” organization of nucleoli, comprising three distinct subcompartments arranged concentrically from the innermost to the outermost layers \cite{boisvert2007multifunctional, feric2016coexisting, lafontaine2021nucleolus}. This nested organization facilitates ribosome biogenesis: ribosome assembly begins in the innermost core and proceeds outward through successive layers for further maturation and modification \cite{boisvert2007multifunctional, feric2016coexisting, lafontaine2021nucleolus, Quinodoz2024}. Another example is the Janus-like configuration formed by the docking of stress granules with P~bodies, where their shared interface enables the dynamic exchange of mRNA and regulatory proteins, thereby coordinating mRNA metabolism and translation \cite{decker2012p, youn2019properties, sanders2020competing}.

\begin{figure*}[!htb]
	\centering
	\includegraphics[width=1\linewidth]{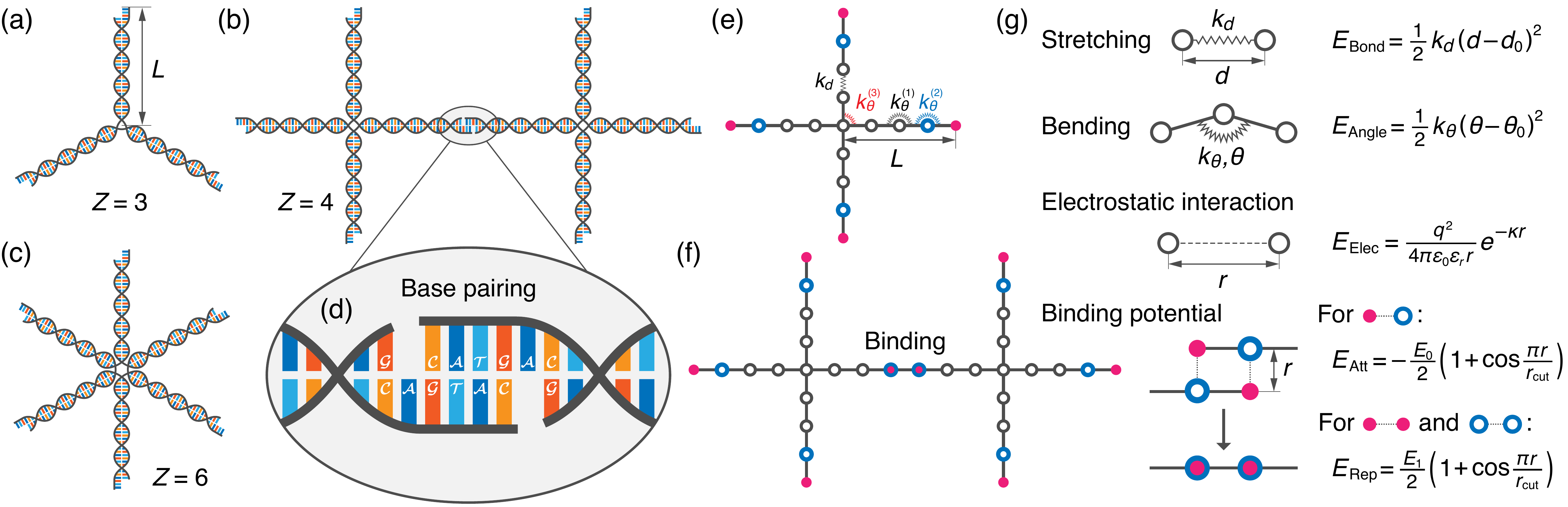}
	\caption{
    Structures of DNA nanostars and their coarse-grained representations.
    Each DNA nanostar consists of a central junction, \(Z\) double-stranded arms of length \(L\), and \(Z\) single-stranded sticky ends.
    (a–c)~DNA nanostars with valence \(Z = 3\), \(4\), and \(6\), respectively.
    (b,~d)~Binding between two DNA nanostars. Base-pairing follows standard rules: adenine (“\(\!\mathcal{A}\)”) pairs with thymine (“\(\mathcal{T}\)”), and cytosine (“\(\mathcal{C}\)”) pairs with guanine (“\(\mathcal{G}\)”). The sticky ends are typically designed to bind only with their own kind, which requires palindromic sequences [the forward strand reads the same as the reverse complement, e.g., “\(\mathcal{G\!T\!\!AC}\)” in (d)]. An additional unpaired base between the sticky end and the double-stranded arm [the adenine in (d)] facilitates reversible unbinding and thus enhances condensate fluidity \cite{jeon2020sequence, nguyen2017tuning}.
    (e–g)~Coarse-grained model of DNA nanostars, incorporating stretching and bending springs as well as non-bonded interactions. Sticky ends are represented by colored beads: blue beads attract red ones, whereas beads of the same color repel each other. See {\it Methods} for modeling details.
    }
	\label{fig1}
\end{figure*}

While continuum theories clarify why specific interfacial energy relations are required to produce different morphologies, the microscopic mechanisms by which cells tune the properties of their constituent molecules to achieve such interfacial energy differences remain unclear. Ultimately, the interfacial energy of a condensate arises from interactions among its constituent molecules, which are governed by a range of molecular features—including molecular size, shape, sequence, valence, interaction strength, and stoichiometry—as well as environmental conditions such as temperature, pressure, surfactant composition, and salt concentration \cite{schuster2021biomolecular, fisher2020tunable, alshareedah2021quantifying, pyo2022surface, laghmach2022rna, joseph2021thermodynamics, kudlay2007sharp, pyo2023proximity}. Because it is impractical to explore all possible molecular architectures, we focus on a specific, experimentally designable class of materials—DNA nanostars \cite{park2009cell, kurokawa2017dna, nishikawa2011biodegradable, um2006enzyme, xing2018microrheology, tanase2025internal, abraham2025cooperation, chaderjian2025diverse}—which are representative of phase-separating biopolymers and serve as ideal model systems for exploring how microscopic parameters shape macroscopic morphologies.

A DNA nanostar is a branched nanostructure in which multiple DNA strands are connected at a central node, and each arm terminates with a single-stranded “sticky end” capable of hybridizing with complementary ends on other nanostars (Fig.~\ref{fig1}a–d). Biomolecular condensates are enriched in multivalent structures, such as the star-like pentamers in nucleoli \cite{mitrea2014structural, mitrea2016nucleophosmin, feric2016coexisting, lafontaine2021nucleolus} and the messenger RNA–protein complexes (mRNPs) in stress granules \cite{buchan2014mrnp, jain2016atpase}. Although multivalent linear chains differ topologically from star polymers, many biopolymers—such as RNA and proteins—can fold or assemble into compact architectures with multiple interaction sites \cite{oldfield2014intrinsically, buchan2014mrnp, mitrea2014structural, mitrea2016nucleophosmin, feric2016coexisting, jain2016atpase, sanders2020competing, xu2020rigidity, lafontaine2021nucleolus, jacobs2023theory}. Such zero-dimensional–like structures can be effectively modeled using DNA nanostars to investigate how biomolecular multivalency influences phase separation and the emergence of diverse morphologies. In particular, because the structure and binding interactions of DNA nanostars can be precisely tuned through sequence design, they provide a versatile bottom-up platform for constructing diverse multiphase morphologies {\it in vitro} \cite{jeon2020sequence, sato2020sequence}. However, systematic and quantitative exploration of morphology-design principles through experiments remains inefficient, owing to the challenges of simultaneously parameterizing the many molecular characteristics involved. Therefore, in this study we employ molecular dynamics simulations to systematically analyze how interfacial energies and multiphase morphologies depend on the microscopic properties of DNA nanostars.

Here, we combine a coarse-grained molecular dynamics model for the dynamics of DNA nanostars with a generalized Flory–Huggins lattice model to quantitatively elucidate how their molecular characteristics govern the phase behavior of DNA condensates. Throughout this work, the {\it components} of the system are defined solely by the types of DNA nanostars, with solvent molecules treated implicitly. The phase diagram and interfacial energy of single-component DNA condensates are first investigated. We then analyze mixtures containing two distinct types of DNA nanostars and show that the interfacial energy between the two dense phases can be tuned by the distribution of sticky-end types on each component. The resulting spectrum of morphologies reveals that Janus-like structures predominate, whereas nested configurations are comparatively rare. Additional strategies for designing target morphologies in DNA-nanostar solutions are also explored. The computational model and theoretical framework developed in this work provide insights into the formation of multiphase biocondensates in cells and establish design principles for engineering complex, functional morphologies in {\it vitro}.

\begin{figure*}[!htb]
	\centering
	\includegraphics[width=0.75\linewidth]{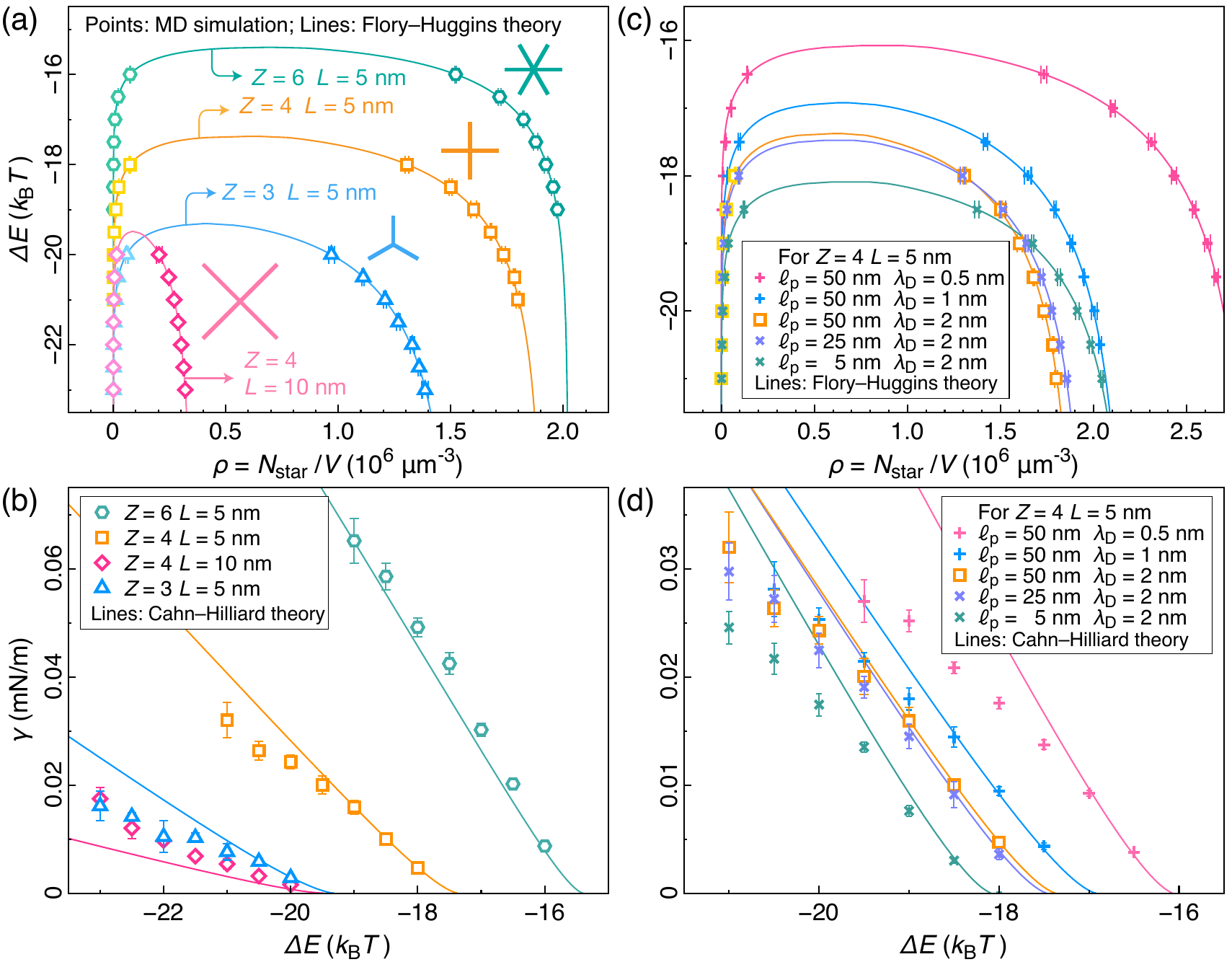}
	\caption{
    (a,~c)~Phase diagrams of systems containing a single type of DNA nanostar with identical sticky ends.
    Here, \(\rho\) is the molecular number density of DNA nanostars in each phase, \(\Delta E\) is the sticky-end binding energy, \(Z\) is the valence, and \(L\) is the arm length. The points represent molecular dynamics (MD) simulation results, and the solid curves show the corresponding Flory–Huggins binodal lines obtained with fitted parameters [\(M_{\rm eff}\) and \(\rho_0\) in Eq.~(\ref{Eq:Flory}); \(z_{\rm eff}\) and \(C\) in Eq.~(\ref{Eq:Florychi})]. Within the coexistence region, the system separates into two phases—a dilute phase and a dense phase—the densities of which correspond to the left and right branches of the binodal, respectively.
    (b, d)~Interfacial energies of different DNA condensates as a function of sticky-end binding energy. The points represent molecular dynamics results, and the solid lines show theoretical predictions from Eq.~(\ref{Eq:florygamma}), where the characteristic interfacial length scale is taken as \(\lambda = \mathcal{C}\rho_0^{-1/3}\). The prefactor \(\mathcal{C}=0.64\) is fitted from the data in (b) and subsequently applied in (d). In (c, d), only nanostars with \(Z=4\) and \(L=5~\rm nm\) are considered, while the persistence length \(\ell_{\rm p}\) and Debye length \(\lambda_{\rm D}\) are varied to examine their effects on interfacial energy [\(\ell_{\rm p}=50~\rm nm\) and \(\lambda_{\rm D}=2~\rm nm\) in (a, b)].
    All data points in (a–d) are obtained from simulations at overall molecular number densities of \(0.42\times10^6~{\rm \mu m}^{-3}\) for \(L=5~\rm nm\) and \(0.052\times10^6~{\rm \mu m}^{-3}\) for \(L=10~\rm nm\), both near their respective critical densities. Error bars represent the standard error of the mean (SEM) from five independent simulations.
    }
    \label{fig2}
\end{figure*}

\section{Results}
\subsection{Coarse-grained model of DNA nanostars}

Numerous coarse-grained models of DNA have been developed \cite{ouldridge2011structural, dans2016multiscale, uusitalo2015martini, de2011coarse, knotts2007coarse}, most of which represent DNA as a complex polymer incorporating structural features such as double strands and base pairing. Such detailed models are not well-suited for simulating large-scale phase separation of DNA nanostars, which involves at least thousands of molecules. To address this, we develop a bead–spring model for DNA nanostars that omits internal base-pairing details. In this coarse-grained model (Fig.~\ref{fig1}e–g), each DNA arm is represented by a bead–spring chain with resistance to both stretching and bending. All arms in a given nanostar are connected to a central bead. At the tip of each arm, two additional beads represent the sticky end (Fig.~\ref{fig1}f,~g): red beads attract blue beads, while beads of the same color repel each other, ensuring that no more than two sticky ends can bind simultaneously. In addition, DNA nanostars repel each other via screened electrostatic interactions: the negatively charged DNA backbones are partially neutralized by counterions in solution, leading to effective screening characterized by the Debye length, \(\lambda_{\rm D}\), which decreases with increasing salt concentration. Molecular dynamics simulations were performed using LAMMPS \cite{plimpton1995fast}. Details of the interaction potentials and simulation procedures are provided in {\it Methods} and Table~S2.

\subsection{Phase diagram and interfacial energy of one-component systems}

We first consider simple systems composed of a single type of DNA nanostar, each possessing identical sticky ends that mutually attract. DNA nanostars with only one kind of sticky end are referred to as {\it homogeneous}. Each DNA nanostar is characterized by its size and valence, quantified by the arm length, \(L\), and the number of arms, \(Z\), respectively (Fig.~\ref{fig1}). As shown in Fig.~\ref{fig2}a, we calculate phase diagrams of the sticky-end binding energy \(\Delta E<0\) versus the molecular number density \(\rho\) for DNA nanostars with different molecular characteristics. The temperature is fixed at 298~K in all simulations, since variations in \(\Delta E\) effectively mimic changes in temperature. The phase diagram exhibits an upper critical binding energy \(\Delta E_{\rm c}\) and a corresponding critical density \(\rho_{\rm c}\); below \(\Delta E_{\rm c}\), a coexistence line appears. Higher valence and smaller nanostar size require a weaker \(|\Delta E_{\rm c}|\) to induce phase separation (or equivalently, correspond to a higher critical temperature at fixed \(\Delta E\)). Increasing valence also yields more compact condensates with higher number densities \(\rho\), as it gives rise to dynamic networks with greater connectivity. The valence-dependent trends in \(\Delta E_{\rm c}\) and \(\rho\) observed here are consistent with previous studies on DNA nanostars \cite{biffi2013phase, locatelli2017condensation, conrad2022emulsion, rovigatti2014accurate} and closely resemble those reported for patchy particles—particles with multiple directional binding sites—in the liquid–gas coexistence region \cite{bianchi2006phase, bianchi2008theoretical}. Such correspondence indicates that DNA nanostars effectively behave as patchy particles, which can be regarded as quasi–zero-dimensional entities owing to their localized interactions and limited spatial extent. This conceptual view underlies the lattice model developed in the next section, which closely reproduces the molecular dynamics results.

Figure~\ref{fig2}b shows the relationship between the binding energy \(\Delta E\) and the interfacial energy of phase-separated condensates, \(\gamma\), calculated using the Kirkwood–Buff formula \cite{kirkwood1949statistical} (see {\it Methods}). The interfacial energy generally increases with \(|\Delta E-\Delta E_{\rm c}|\); for a fixed \(\Delta E\), nanostars with higher valence or smaller size exhibit higher interfacial energies. Doubling the size of DNA nanostars markedly reduces \(\gamma\), as larger molecules have fewer binding sites per unit volume and greater structural flexibility. To examine the effect of arm flexibility more clearly, we fix the DNA arm length at \(L = 5\)~nm and vary the bending rigidity of the arms. As shown in Fig.~\ref{fig2}c,~d, decreasing the bending rigidity—equivalently, reducing the persistence length \(\ell_{\rm p}\)—increases the critical binding strength \(|\Delta E_{\rm c}|\) required for phase separation. For a fixed binding energy, the reduced persistence length lowers \(\gamma\), corresponding to a leftward shift of the interfacial-energy curve in Fig.~\ref{fig2}d. These changes arise from stronger thermal fluctuations in more flexible nanostars, which increase the entropic contribution to the effective free energy between two nanostars, thereby weakening binding and lowering the interfacial energy at constant \(\Delta E\). We also vary the Debye screening length \(\lambda_{\rm D}\) (Fig.~\ref{fig2}c,~d) and find that shorter screening lengths yield smaller \(|\Delta E_{\rm c}|\) and produce rightward shifts of the interfacial-energy curve. This occurs because shorter \(\lambda_{\rm D}\) promotes molecular encounters and binding, thereby facilitating phase separation and yielding denser condensates at the same \(\Delta E\) (Fig.~\ref{fig2}c). In addition, near the critical point \((\Delta E_{\rm c}, \rho_{\rm c})\), thermodynamic quantities such as the interfacial energy follow a power-law relation with a universal exponent independent of microscopic details \cite{hohenberg1977theory}. Specifically, \(\gamma \propto |\Delta E - \Delta E_{\rm c}|^{\mu}\), where \(\mu \approx 1.26\) for the three-dimensional Ising universality class \cite{widom1965surface, chaikin1995principles, pyo2023proximity}. The results in Fig.~\ref{fig2}b,~d are consistent with the predicted power-law scaling as \(\Delta E\) approaches the critical point (see Fig.~S6). In the opposite limit, where \(|\Delta E - \Delta E_{\rm c}|\) is large, strong binding drives the formation of a solid-like gel phase with negligible fluidity on the simulation timescale.

\subsection{Flory–Huggins theory and generalized lattice model}

\begin{figure}[!t]
    \centering
    \includegraphics[width=0.7\linewidth]{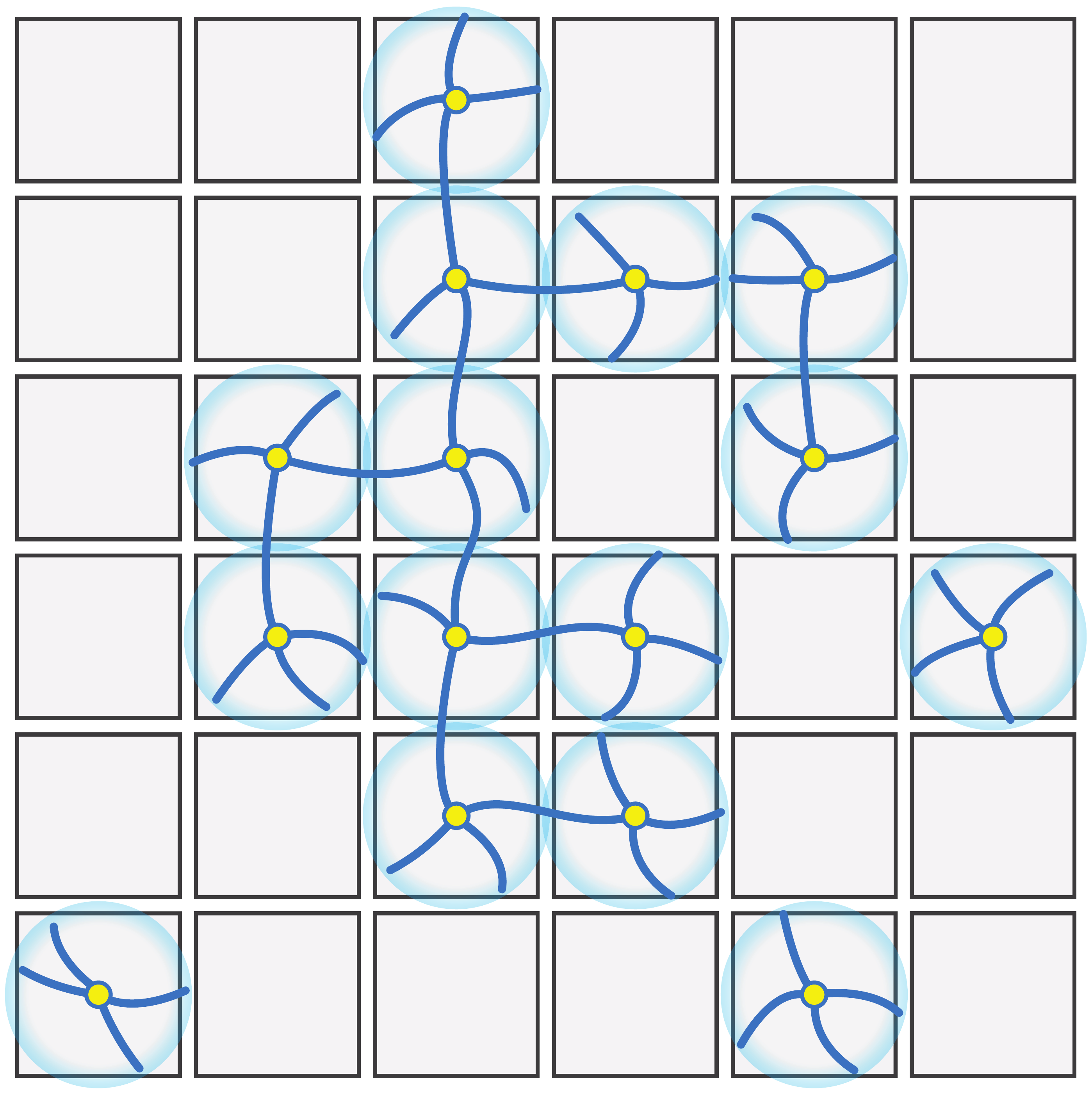}
    \caption{Schematic of the lattice model for DNA-nanostar solutions. Each nanostar occupies one lattice site and interacts with its nearest neighbors. Unoccupied sites are filled with pure solvent, while each occupied site also includes solvent surrounding the nanostar. The central junction of each nanostar (yellow node) is located at the site center, and its sticky ends are distributed within a blue spherical shell that partially overlaps with those of neighboring sites.
    }
    \label{fig2.4}
\end{figure}

To elucidate how the molecular parameters of DNA nanostars govern the phase behavior observed in Fig.~\ref{fig2}, we develop a Flory–Huggins model consistent with our molecular dynamics simulations and capable of describing multicomponent systems. A brief overview of the Flory–Huggins framework for DNA-nanostar solutions is presented below, with detailed derivations in Appendix~A of {\it Supporting Information}. Following the classical Flory–Huggins treatment of polymer solutions \cite{flory1942thermodynamics, huggins1941solutions}, we construct a generalized lattice model in which each DNA nanostar is fixed at the center of a lattice cell (Fig.~\ref{fig2.4}). Each occupied lattice site contains one DNA nanostar and its surrounding solvent molecules, while the unoccupied sites are filled entirely with solvent. The arms of each nanostar can rotate around the central node and interact either sterically with other arms on the same molecule or through binding with arms from neighboring molecules. Binding occurs only between arms on adjacent lattice sites, since the compact geometry of DNA nanostars renders intramolecular binding energetically unfavorable due to bending penalties. Unlike conventional lattice models, in which neighboring occupied sites automatically contribute a favorable interaction energy, adjacent DNA nanostars in our model do not necessarily lower the energy unless their sticky ends actually bind. The total number of lattice sites is \(N\), and the number of nanostar molecules is \(N_{\rm star}\). The free energy of mixing, \(F\), is thus given by \cite{flory1942thermodynamics, huggins1941solutions}
\begin{align} \label{Eq:Flory}
    \frac{F}{Nk_{\rm B}T} =
    \frac{\phi}{M_{\rm eff}} \ln \phi
    + (1-\phi) \ln (1-\phi)
    + \chi \phi (1-\phi),
\end{align}
where \(k_{\rm B}\) is the Boltzmann constant, \(T\) the temperature, \(\phi\) the volume fraction of lattice sites occupied by DNA nanostars, and \(\chi\) the Flory interaction parameter. The volume fraction is defined as \(\phi \equiv N_{\rm star}/N = \rho_{\rm star}/\rho_0\), where \(\rho_{\rm star} = N_{\rm star}/V\) and \(\rho_0 = N/V\) denote the number densities of DNA nanostars and lattice sites, respectively, and \(V\) is the total system volume. Note that \(\phi\) represents the fraction of occupied lattice sites rather than the actual volume fraction of DNA nanostars—the system still contains a substantial amount of solvent even when \(\phi = 1\). Consequently, the lattice number density \(\rho_0\) determines the effective width of the binodal region in Fig.~\ref{fig2}, which varies among different DNA nanostars. In Eq.~(\ref{Eq:Flory}), the parameter \(M_{\rm eff} > 1\) serves as a correction factor that accounts for the asymmetry of the binodal lines in Fig.~\ref{fig2}. This behavior resembles that of polymer solutions \cite{flory1942thermodynamics, huggins1941solutions} and suggests that DNA nanostars effectively form larger polymer-like assemblies, with \(M_{\rm eff}\) representing an average polymerization factor.

The Flory interaction parameter \(\chi\) in Eq.~(\ref{Eq:Flory}) characterizes, at a mean-field level, how molecular interactions govern the miscibility between occupied and unoccupied lattice sites, effectively incorporating all microscopic details of the system. It serves as a key parameter controlling the phase behavior of DNA-nanostar solutions. In the generalized lattice model, neighboring nanostars can either bind or remain unbound, resulting in a change in conformational entropy upon binding. The Flory interaction parameter \(\chi\) is approximated from the effective free energy per pair of neighboring nanostars, obtained after integrating out all microscopic configurations, and is expressed as
\begin{align} \label{Eq:Florychi}
    \chi \approx \frac{z_{\rm eff}-2}{2}
    \left[
    -\frac{\Delta E}{k_{\rm B}T}
    + 2 \ln Z
    - 3 \ln L 
    + C (\lambda_{\rm D})
    \right].
\end{align}
Here, \(z_{\rm eff}\) denotes the effective coordination number of the generalized lattice, \(\Delta E\) the sticky-end binding energy, \(Z\) and \(L\) the valence and arm length of DNA nanostars, respectively, and \(C (\lambda_{\rm D})\) a monotonically decreasing function of the Debye screening length \(\lambda_{\rm D}\) (see Appendix~A). In Eq.~(\ref{Eq:Florychi}), the \(\Delta E\) term represents the energetic contribution to the effective free energy, whereas the remaining terms arise from entropic effects associated with molecular configurations and accessible binding states. For instance, the term \(2\ln Z\) accounts for the combinatorial entropy of binding between two nanostars with valence \(Z\). In contrast, the term \(-3\ln L\) represents the entropy loss upon binding, since the paired sticky ends have greatly reduced spatial freedom. This loss becomes more pronounced for larger \(L\), where the number density of sticky ends is lower. Note that \(\chi\) does not explicitly depend on the arm persistence length \(\ell_{\rm p}\), because Eq.~(\ref{Eq:Florychi}) assumes \(L \ll \ell_{\rm p}\), i.e., that each DNA arm behaves as a nearly rigid segment. The predictions based on this assumption are consistent with the simulation results in Fig.~\ref{fig2}c,~d: the phase-diagram and interfacial-energy curves remain nearly unchanged between \(\ell_{\rm p}=50\)~nm and 25~nm, but deviate markedly once \(\ell_{\rm p}\) becomes equal to \(L=5\)~nm. In practice, most experimentally synthesized DNA nanostars have short arms (\(L \lesssim 10\ \rm nm\)) compared with the DNA persistence length (\(\ell_{\rm p} \approx 50\)~nm) \cite{jeon2020sequence, sato2020sequence}, validating the applicability of Eq.~(\ref{Eq:Florychi}). As the critical Flory interaction parameter \(\chi_{\rm c}\) and the coordination number are similar across different DNA nanostars, Eq.~(\ref{Eq:Florychi}) directly explains how valence, arm length, and Debye length determine the critical binding energy \(\Delta E_{\rm c}\), consistent with the trends in Fig.~\ref{fig2}.

To extend the Flory–Huggins framework to multicomponent systems, we consider \(n\) distinct types of DNA nanostars together with solvent, represented collectively as \(n+1\) species in the lattice model. The free-energy density \(f\), including interfacial contributions, is given by
\begin{align} \label{Eq:fx}
    \!\!\!\frac{f}{\rho_0 k_{\rm B}T} =
    \sum_{i=0}^n
    \frac{\phi_i}{\!M_{i,{\rm eff}}\!} \ln \phi_i
    +
    \frac{1}{2} \!
    \sum_{i,j=0}^n\!
    \chi_{ij} \!\left(
    \phi_i\phi_j
    \!-\!
    \lambda^2 \nabla \phi_i \!\cdot\! \nabla \phi_j
    \right)\!.\!\!\!
\end{align}
Here, \(\phi_i\) denotes the volume fraction of species \(i\): for \(i>0\), occupied lattice sites containing DNA nanostars of type \(i\); and for \(i=0\), solvent-filled sites, satisfying \(\sum_{i=0}^n \phi_i = 1\). The factor \(M_{i,{\rm eff}}\) represents the effective polymerization of species \(i\), with \(M_{0,{\rm eff}} = 1\) for monomeric solvent sites. The parameter \(\chi_{ij}\) denotes the Flory interaction between distinct species \(i\) and \(j\) (\(\chi_{ii}=0\)). The characteristic length \(\lambda\) sets the interfacial width, and the gradient terms describe interfacial-energy contributions. For simplicity, the lattice-site number density \(\rho_0\) is assumed constant across all types of nanostars, implying comparable effective molecular sizes.

The Flory interaction parameter \(\chi_{i0}\) between a DNA nanostar of type \(i\) (\(i\neq 0\)) and the solvent retains the form of Eq.~(\ref{Eq:Florychi}), but we now generalize it to the case where each nanostar possesses two distinct types of sticky ends, A and B (termed {\it heterogeneous}), which is necessary to generate distinct interfaces and, consequently, diverse condensate morphologies. The sticky ends can be designed by altering their nucleotide sequences (e.g., “\(\mathcal{CGCGCG}\)” versus “\(\mathcal{\!AT\!\!AT\!\!AT}\)”). Here, we specifically consider systems in which all sticky-end pairs (A–A and B–B) share the same binding energy \(\Delta E_{\rm AA} = \Delta E_{\rm BB} = \Delta E\), whereas different types of sticky ends interact repulsively. The general case with \(\Delta E_{\rm AA} \neq \Delta E_{\rm BB}\) is discussed in {\it Supporting Information}. A nanostar of type \(i\) is defined as “\(\alpha_i{\rm A}\beta_i{\rm B}\)” if it contains \(\alpha_i\) sticky ends of type A and \(\beta_i\) sticky ends of type B. The generalized Flory interaction parameter \(\chi_{i0}\) for \(i \neq 0\) is then given by
\begin{align} \label{Eq:Florychii}
    \!\chi_i \equiv \chi_{i0}
    &\approx \frac{z_{i, \rm eff}\!-\!2}{2}
    \!\left[
    -\frac{\Delta E}{k_{\rm B}T}
    +
    \ln (\alpha_{i}^2 
    \!+\!
    \beta_{i}^2)
    \!-
    3 \ln L_i + C(\lambda_D)
    \right]\!,\!\!
\end{align}
where \(z_{i,{\rm eff}}\) denotes the effective coordination number of the lattice sites occupied by nanostars of type \(i\). The term \(\ln(\alpha_{i}^2+\beta_{i}^2)\) represents the combinatorial entropy associated with pairing two identical heterogeneous nanostars, which reduces to \(2 \ln Z_i\) in the homogeneous limit [$\alpha_i=Z_i, \beta_i=0$, see Eq.~(\ref{Eq:Florychi})]. The Flory interaction parameter \(\chi_{ij}\) between two nanostar types (\(i,j \neq 0\)) is given by
\begin{align} \label{Eq:Florychiij}
    \!\!\chi_{ij} \approx 
    \frac{z_{i,\rm eff}\!-\!2}{2} \ln\frac{\alpha_i^2+\beta_i^2}{\alpha_i\alpha_j + \beta_i\beta_j}
    + \frac{z_{j,\rm eff}\!-\!2}{2} \ln\frac{\alpha_j^2+\beta_j^2}{\alpha_i\alpha_j + \beta_i\beta_j},\!
\end{align}
where we have assumed that the arm length \(L_i\) is identical across all nanostar types, consistent with the condition used in the following section. Equation~(\ref{Eq:Florychiij}) automatically satisfies \(\chi_{ii}=0\) and is derived under the condition \(\alpha_i\alpha_j + \beta_i\beta_j \neq 0\), which indicates that a nanostar of type \(i\) can bind to a nanostar of type \(j\). Otherwise, the two nanostars are termed “orthogonal,” with \(\chi_{ij} \approx \chi_i + \chi_j\), as discussed in {\it Supporting Information}. Note that Eq.~(\ref{Eq:Florychiij}) indicates that the \(n\)-component Flory–Huggins model requires no additional input beyond that of the \(n\)~corresponding single-component systems, since parameters such as \(z_{i,{\rm eff}}\) and \(M_{i,{\rm eff}}\) can be determined from mixtures containing only component \(i\) and solvent. Equations~(\ref{Eq:Florychii}) and~(\ref{Eq:Florychiij}) can be further generalized to DNA nanostars with arbitrary sticky-end distributions (e.g., “1A2B2C1D”), with the combinatorial terms modified accordingly.

Given all Flory interaction parameters \(\chi_{ij}\), the interfacial energy between two equilibrium phases \(I\) and \(J\) is evaluated based on the Cahn–Hilliard formalism \cite{cahn1958free} (see Appendix~B of {\it Supporting Information} for full derivation):
\begin{align} \label{Eq:florygamma}
    \gamma_{I\!J}
    = 2 \lambda \rho_0 k_{\rm B}T \int_{0}^{1}
    {\rm d}\eta
    \sqrt{K_{I\!J} \, \Delta\overline f(\eta)}.
\end{align}
Here, \(\eta \in [0,1]\) parameterizes the local composition across the interface, with \(\phi_i(\eta) = \eta \phi_i^I + (1-\eta)\phi_i^J\), and \(\phi_i^I\) denotes the equilibrium volume fraction of species~\(i\) in phase~\(I\).
The effective coupling \(K_{I\!J}\) between the two phases is given by
\begin{align}
    K_{I\!J} =
    -\frac{1}{2} \sum_{i, j=0}^n \chi_{ij} (\phi_i^I - \phi_i^J) (\phi_j^I - \phi_j^J),
\end{align}
and the excess dimensionless free energy associated with the non-equilibrium composition at the interface is
\begin{align} \label{Eq:fexcess}
    \Delta \overline f(\eta) =
    \overline f(\{\phi_i(\eta)\})
    - \eta \overline f(\{\phi_i^I\})
    - (1-\eta) \overline f(\{\phi_i^J\}),
\end{align}
where the homogeneous-state free energy takes the form
\begin{align} \label{Eq:fhomo}
    \overline f (\{\phi_i\}) =
    \sum_{i=0}^n \frac{\phi_i}{M_{i, {\rm eff}}} \ln \phi_i
    +
    \frac{1}{2} 
    \sum_{i,j=0}^n
    \chi_{ij} 
    \phi_i\phi_j.
\end{align}
In the limit of \(\chi_{ij}\gg 1\) for all \(i\)~and~\(j\), we can label phases~\(I\) and~\(J\) such that they are predominantly composed of species~\(I\) and~\(J\), respectively, i.e., \(\phi_i^I\approx\delta_{iI}\) and \(\phi_i^J\approx\delta_{iJ}\), with \(\delta_{ij}\) the Kronecker delta. In this case, the homogeneous free energy in Eq.~(\ref{Eq:fhomo}) is dominated by the second (interaction) term, and the excess free energy in Eq.~(\ref{Eq:fexcess}) can then be approximated as \(\Delta \overline f \approx \eta(1-\eta) \chi_{I\!J}\). Since the effective coupling term likewise satisfies \(K_{I\!J} \approx \chi_{I\!J}\), the interfacial energy is thus given by~\cite{mao2019phase}
\begin{align} \label{Eq:gamma_approx}
    \gamma_{I\!J}
    \approx \frac{\pi}{4}\lambda \rho_0 k_{\rm B}T \chi_{I\!J}
    .
\end{align}
As shown in Fig.~\ref{fig2}, the coexistence line and the interfacial energy of each single-component DNA condensate are calculated from the Flory–Huggins theory, showing good agreement with molecular dynamics simulations. Equations~(\ref{Eq:Florychi}) and (\ref{Eq:gamma_approx}) indicate that the slope of each interfacial-energy curve in Fig.~\ref{fig2}b,~d is proportional to \(\lambda\rho_0(z_{\rm eff}-2)\) at stronger binding energy \(|\Delta E|\), where the lattice-site density \(\rho_0\) and lattice coordination number \(z_{\rm eff}\) can be determined by fitting the theoretical binodal to simulation results. The characteristic length \(\lambda\), which sets the scale of the interfacial width, is assumed proportional to the lattice spacing in the generalized lattice model, given by \(\lambda = \mathcal{C}\rho_0^{-1/3}\). The factor \(\mathcal{C}\) may vary across systems as the coordination number of the generalized lattice changes. However, a single value of \(\mathcal{C}=0.64\) is used for all cases, accounting for the small discrepancies between theory and simulation results.

\subsection{Interfacial energies in two-component systems}

\begin{figure*}[!t]
	\centering
    \begin{minipage}[t]{0.5\linewidth}
	\includegraphics[width=1\linewidth]{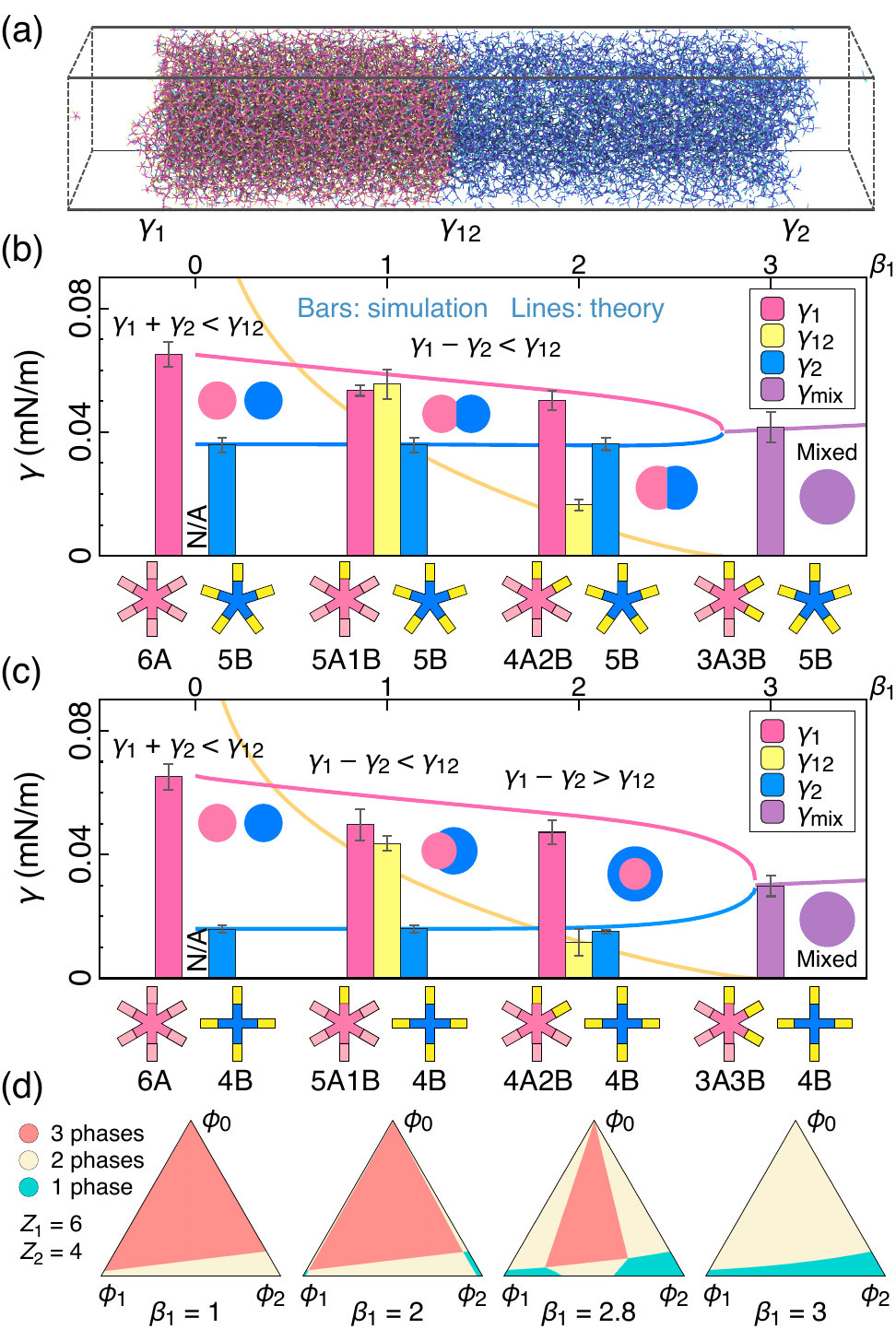}
    \end{minipage}%
    \hfill
    \begin{minipage}[t]{0.45\linewidth}
    \vspace{-13.4 cm}
	\caption{
    (a)~Simulation snapshot of a two-component system of DNA nanostars used to compute interfacial energies. Here, \(\gamma_1\) and \(\gamma_2\) denote the interfacial energies between the dilute phase and the pink or blue dense phase, respectively, while \(\gamma_{12}\) represents the interfacial energy between the pink and blue phases. The simulation box measures \(1000\times100\times100~{\rm nm}^3\) [only a portion along the \(x\)-direction is shown] and contains 4000 pink nanostars (sticky-end distribution 4A2B) and 4000 blue nanostars (4B). Periodic boundary conditions are applied in all directions.
    (b,~c)~Interfacial energies for different combinations of DNA nanostars. Pink and blue bars correspond to \(\gamma_i\) between dense phase~\(i\) and the dilute phase; yellow bars represent \(\gamma_{12}\) between the two dense phases [see (a)]; and purple bars denote the interfacial energy of the mixed dense phase, \(\gamma_{\rm mix}\), where the two components coexist within a single phase. All bar values are obtained from molecular dynamics simulations. Solid lines show theoretical predictions based on Eqs.~(\ref{Eq:Florychii}) and~(\ref{Eq:Florychiij}), plotted as a function of \(\beta_1\) (the number of yellow sticky ends on the pink nanostar) indicated on the upper horizontal axis. The insets in (b,~c) depict the morphologies predicted from interfacial-energy inequalities informed by molecular dynamics calculations. In the mixed-phase regime, \(\gamma_1\), \(\gamma_2\), and \(\gamma_{12}\) are undefined, and only \(\gamma_{\rm mix}\) applies. The binding energy of identical sticky ends is fixed at \(\Delta E=-19k_{\rm B}T\), and the DNA arm length is \(L=5~\rm nm\). Distinct sticky ends repel each other. Error bars indicate the standard error of the mean (SEM) from five independent simulations.
    (d)~Representative ternary phase diagrams corresponding to the systems in (c), showing variations in \(\beta_1\). Here, \(\phi_0\), \(\phi_1\), and \(\phi_2\) denote the volume fractions of lattice sites occupied by the solvent, component~1, and component~2, respectively.
    }
	\label{fig3}
    \end{minipage}
\end{figure*}

With the Flory–Huggins framework for DNA nanostars established, we now focus on systems composed of two distinct types of DNA nanostars bearing sticky ends A and B, denoted “\(\alpha_i{\rm A}\beta_i{\rm B}\)” for \(i\in\{1,2\}\). The binding energies are chosen as \(\Delta E = \Delta E_{\rm AA} = \Delta E_{\rm BB} = -19k_{\rm B}T\), with only steric repulsion between A and B. Importantly, A and B denote only the sticky-end segments; the remaining portions of the DNA arms still interact via electrostatic repulsion. The choice of \(\Delta E\) is guided by Fig.~\ref{fig2}b, where condensates formed by four-arm nanostars exhibit relatively small interfacial energies. If \(\Delta E\) were more negative (stronger binding), the nanostars would form a solid-like gel phase that lacks fluidity on the simulation timescale. All nanostars are assigned a fixed arm length \(L=5\)~nm, persistence length \(\ell_{\rm p} = 50\)~nm, and Debye length \(\lambda_{\rm D} = 2\)~nm; only their valence and sticky-end distributions are varied. For each sticky-end distribution—for example, 4A2B—we only consider the configuration “AAAABB,” while the other possible arrangements, “AAABAB” and “AABAAB,” are neglected. These variants yield nearly identical interfacial energies (see Fig.~S3), because the node of a DNA nanostar is semiflexible: in our coarse-grained model, nodal rigidity is assigned only between adjacent arms.

As shown in Fig.~\ref{fig3}, the system consists of two types of DNA nanostars: a pink species bearing sticky ends~A and~B, and a blue species possessing only sticky ends~B. Figure~\ref{fig3}a shows a molecular dynamics snapshot used to compute interfacial energies between coexisting phases. A slab geometry (an elongated rectangular simulation box) is employed to eliminate Laplace-pressure effects from curved interfaces. The interfacial energies obtained from molecular dynamics are evaluated using a regional form of the Kirkwood–Buff formula \cite{kirkwood1949statistical} (see {\it Methods}), and the results for different component combinations are shown as colored bars in Fig.~\ref{fig3}b,~c. Specifically, pink and blue bars correspond to the interfacial energies between the dilute phase and the pink or blue condensates (composed primarily of their respective nanostars), \(\gamma_i \equiv \gamma_{i0}\) (\(i \in \{1,2\}\)); yellow bars indicate the interface between the pink and blue phases, \(\gamma_{12}\); and purple bars represent the interface between the mixed (pink–blue) phase and the dilute phase, \(\gamma_{\rm mix}\). Solid lines show the theoretical interfacial energies predicted by the Flory–Huggins model [see Eq.~(\ref{Eq:florygamma})]. Insets in Fig.~\ref{fig3}b,~c illustrate the morphologies inferred from the relative magnitudes of the interfacial energies obtained from molecular dynamics simulations. For example, if the interfacial energies satisfy the triangle inequalities—\(\gamma_{12} + \gamma_{2} > \gamma_{1}\), \(\gamma_{12} + \gamma_{1} > \gamma_{2}\), and \(\gamma_{1} + \gamma_{2} > \gamma_{12}\)—the three phases coexist at a triple junction in the cross-section. Conversely, if \(\gamma_{12} + \gamma_{2} < \gamma_{1}\), phases~1 and~0 do not make contact, and phase~2 occupies the space between them \cite{mao2019phase, mao2020designing}. Figure~\ref{fig3}d presents representative ternary phase diagrams predicted by the Flory–Huggins theory, illustrating how \(\beta_1\), the number of sticky ends~B on the pink nanostars, influences the phase behavior.

Various combinations of a 6-arm DNA nanostar with homogeneous nanostars of valence~5 or~4 are shown in Fig.~\ref{fig3}b and~c, respectively. When the 6-arm nanostars carry only sticky ends~A and the other component carries sticky ends~B, the two condensates remain separated by the dilute phase (see Fig.~S7), owing to strong Coulombic repulsion between them. When one of the sticky ends~A on the 6-arm stars is replaced by~B, the two condensates begin to wet each other (similar to Fig.~\ref{fig3}a), as intercomponent binding becomes available. In this case, the interfacial energy \(\gamma_{12}\) remains relatively high but is still smaller than \(\gamma_1+\gamma_2\), indicating that interface formation is energetically favorable. As expected, \(\gamma_{12}\) decreases further when two sticky ends~A are replaced by~B, as more interfacial binding sites lower the interfacial energy. When the 6-arm nanostars possess an equal distribution of sticky ends~A and~B (3A3B), the two condensates merge into a single dense phase (see Fig.~S7), characterized by an interfacial energy with the dilute phase, \(\gamma_{\rm mix}\). This transition is captured by the ternary phase diagrams in Fig.~\ref{fig3}d, indicating that the effective Flory interaction parameter \(\chi_{12}\) drops below the critical value for demixing when the attractions between the two components become sufficiently strong. For 6-arm and 5-arm combinations (Fig.~\ref{fig3}b), the interfacial energies satisfy the triangular inequality \(\gamma_1 < \gamma_2 + \gamma_{12}\), indicating that the two dense phases adhere and form a shared interface, giving rise to Janus-like morphologies. In contrast, for the 6-arm/4-arm system shown in Fig.~\ref{fig3}c, the larger disparity between \(\gamma_1\) and \(\gamma_2\) leads to the opposite inequality, \(\gamma_1>\gamma_2+\gamma_{12}\), for the 4A2B–4B combination, thereby favoring nested morphologies.

A few additional remarks are worth noting. First, in Fig.~\ref{fig3}b and~c, the interfacial energy \(\gamma_1\) decreases as the sticky-end distribution of 6-arm nanostars changes from 6A to 4A2B. As illustrated in Eq.~(\ref{Eq:Florychii}), we assume that only one pair of sticky ends can bind between two nanostars (see {\it Supporting Information} for validation of this assumption). Under this condition, there are \(6^2=36\) possible binding configurations for two 6A nanostars, compared with \(4^2+2^2=20\) for two 4A2B nanostars. This reduction in entropy leads to a smaller \(\chi_1\) and, consequently, a lower \(\gamma_1\). In contrast, the interfacial energy \(\gamma_2\) remains nearly unchanged when the sticky-end distribution of component~1 is varied, as phase~2—composed predominantly of component~2—is governed by a constant \(\chi_2\). Theoretical results further show that the interfacial energy of the mixed condensate, \(\gamma_{\rm mix}\), increases with the number of sticky ends~B on the 6-arm nanostars (\(\beta_1\)), because the reduced \(\chi_{12}\) [Eq.~(\ref{Eq:Florychiij})] enhances the cohesion within the mixed condensate. Notably, \(\gamma_{\rm mix}\) also depends on the composition ratio of the two nanostar types, approaching \(\gamma_1\) and \(\gamma_2\) in the limits of \(\phi_1/\phi_2 \to \infty\) and \(\phi_1/\phi_2 \to 0\), respectively. In Fig.~\ref{fig3}c, the theoretical value of \(\gamma_{\rm mix}\) is evaluated at the critical mixing ratio, corresponding to the {\it unique} “tie line” in Fig.~\ref{fig3}d that emerges when the three-phase region disappears (for \(\beta_1\) between 2.8 and~3). This critical ratio satisfies \(\phi_1 \approx \phi_2\), consistent with the molecular dynamics setup, thereby yielding good agreement between theoretical and simulation results for \(\gamma_{\rm mix}\).

Noticeable discrepancies arise between simulation and theory for \(\gamma_{12}\) and \(\gamma_1\) when \(\beta_1 = 1\) or~2. The deviation in \(\gamma_{12}\) arises because a uniform interfacial length scale, \(\lambda=\mathcal{C}\rho_0^{-1/3}\), is assumed globally with a fixed \(\mathcal{C}=0.64\) throughout this work. The discrepancy in Fig.~\ref{fig3} suggests that the effective interfacial width between the two dense phases may be slightly larger than that between a dense and a dilute phase. Accordingly, a more refined treatment could replace the term \(\lambda^2 \nabla\phi_i \cdot \nabla\phi_j\) with \(\lambda_{ij}^2 \nabla\phi_i \cdot \nabla\phi_j\) in Eq.~(\ref{Eq:fx}). The deviation in \(\gamma_1\) originates from the assumption in Eq.~(\ref{Eq:florygamma}) that the interfacial profile follows a straight path in the composition space \(\{\phi_i\}\) (\(i\in\{0,1,2\}\)), i.e., \(\phi_i(\eta) = \eta \phi_i^I + (1-\eta)\phi_i^J\). However, in the slab-shaped simulation box, component~2 tends to accumulate at the interface between phases~1 and~0, acting as a surfactant and causing the actual compositional path to deviate from the straight line—an energetically favorable configuration that lowers \(\gamma_{1}\) (see Fig.~S4; we emphasize that this effect solely contributes to the theoretical overestimation of \(\gamma_{1}\), and thus cannot explain the underestimated \(\gamma_{12}\)). In particular, for the 4A2B–4B case (Fig.~\ref{fig3}c), the equilibrium path between phases~1 and~0 would, in principle, yield \(\gamma_1 = \gamma_2 + \gamma_{12}\), since the interface between phases~1 and~0 is intrinsically unstable and tends to be replaced by phase~2. This argument is validated in Fig.~S4 through diffusive phase-field simulations, which reach equilibrium much faster. However, within the timescales accessible to molecular dynamics, this interfacial restructuring is incomplete. As a result, a metastable interface between phase~1 and the dilute phase persists, with only minimal accumulation of component~2. Consequently, the interfacial energy \(\gamma_1\) measured in molecular dynamics reflects a non-equilibrium value that may relax further over longer timescales. To rigorously validate the morphology predictions based on interfacial-energy inequalities, large-scale simulations are required to directly capture the equilibrium morphologies.

\begin{figure*}[!htb]
	\centering
	\includegraphics[width=1\linewidth]{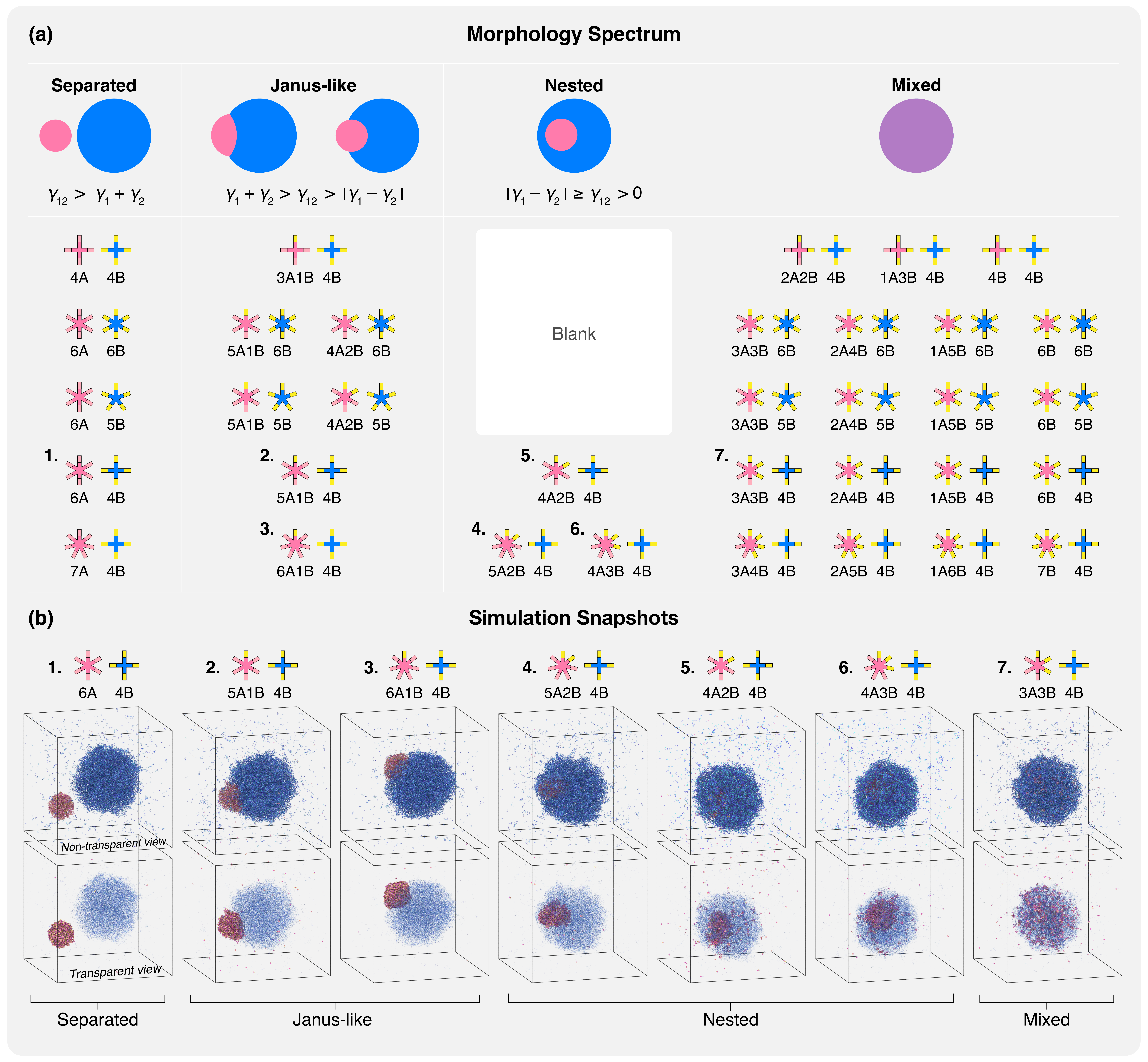}
	\caption{
    (a)~Morphology spectrum for two-component systems of DNA nanostars.
    (b)~Snapshots from molecular dynamics simulations of representative systems. The second row shows the same configurations as the first, but with adjusted transparency to reveal the internal structures. The binding energy of identical sticky ends is fixed at \(\Delta E = -19k_{\rm B}T\), and the arm length is \(L = 5~\rm nm\) for all nanostars. Different types of sticky ends interact purely repulsively. The simulation box measures \(500^3~\rm nm^3\) and contains 1500 pink nanostars and 20000 blue nanostars to clearly visualize the engulfment morphology. Each simulation is run for \(4\times10^8\)~steps, and the corresponding time evolution is shown in Fig.~S8.
}
	\label{fig4}
\end{figure*}

\begin{figure*}[!t]
    \centering
    \includegraphics[width=0.85\linewidth]{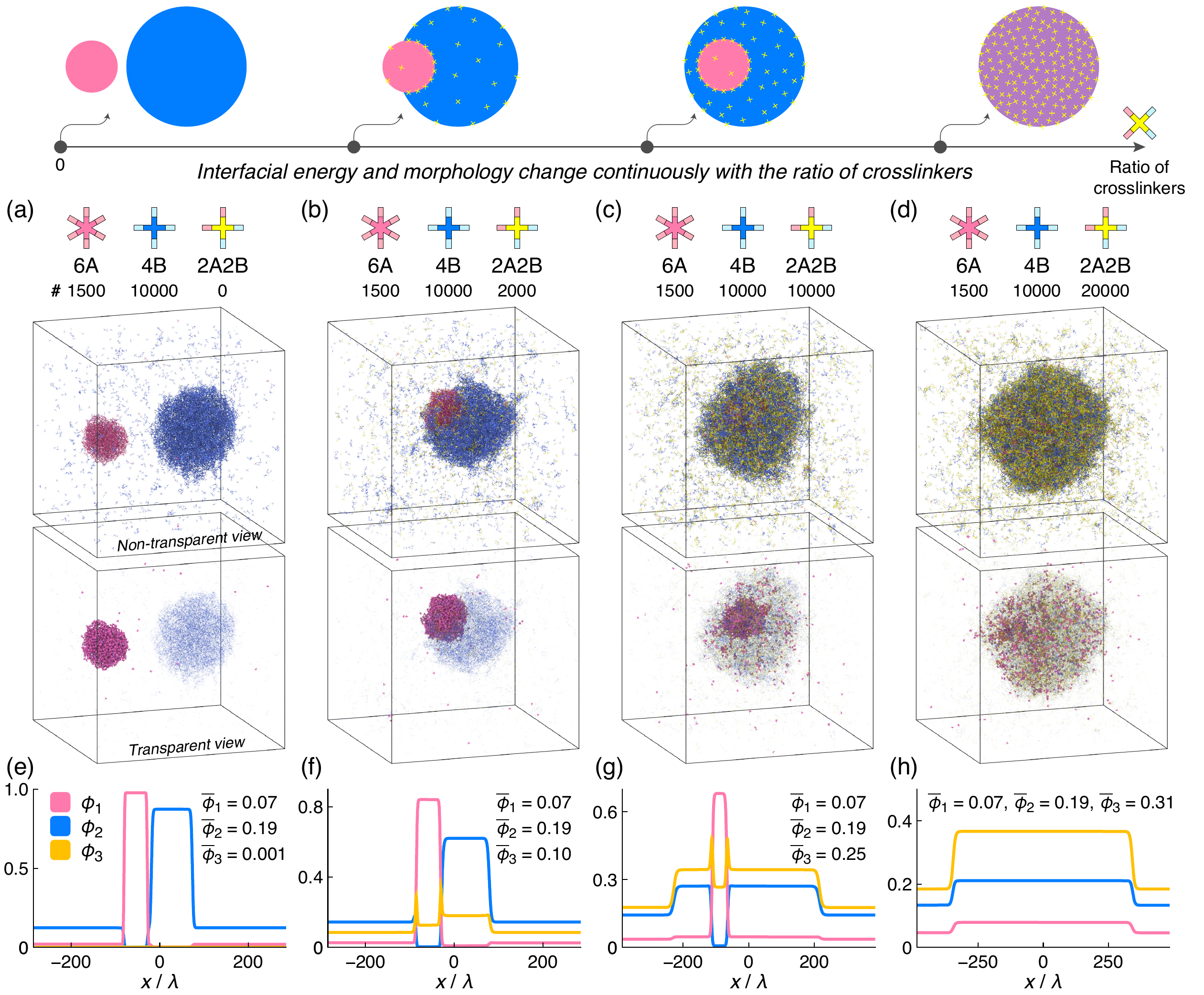}
    \caption{
    (a–d)~Snapshots from molecular dynamics simulations illustrating the formation of different morphologies using crosslinker DNA nanostars. The second row shows the same configurations as the top row, rendered with reduced opacity to reveal the internal structures. The binding energy between identical sticky ends is fixed at \(\Delta E = -19k_{\rm B}T\), and the arm length is \(L = 5~\rm nm\) for all nanostar types. Distinct sticky-end types interact purely repulsively. Each simulation uses a \(500^3~\rm nm^3\) box containing 1500 pink nanostars (6A) and 10000 blue nanostars (4B), while the number of crosslinker nanostars (2A2B) varies from 0 to 20000 across panels (a–d). Each simulation is run independently for \(3\times10^8\)~steps; corresponding time-evolution sequences are shown in Fig.~S9.
    (e–h)~Results from one-dimensional diffusive phase-field (Flory–Huggins) simulations, showing typical equilibrium volume-fraction profiles of each component. The system is periodic with a total length of \(1000\lambda\), where \(\lambda\) is the characteristic length scale associated with the interfacial width [Eq.(\ref{Eq:fx})]. Here, \(\phi_0\), \(\phi_1\), \(\phi_2\), and \(\phi_3\) represent the volume fractions of the solvent and components~1–3, respectively. The crosslinker distribution (\(\phi_3\)) is well resolved within the bulk phases and across the interfaces. The overall volume fraction of each component is denoted by \(\overline{\phi_i}\), with only \(\overline{\phi_3}\) varied to mimic changes in crosslinker concentration. The Flory interaction parameters are set to \(\chi_1 = 4\), \(\chi_2 = 2.6\), \(\chi_3=2.35\), \(\chi_{12} = 6.6\), \(\chi_{13} = 1.46\), and \(\chi_{23} = 1.26\). See Appendix~C in {\it Supporting Information} for details.
    }
    \label{fig4.5}
\end{figure*}

\begin{figure*}[!t]
    \centering
    \includegraphics[width=0.8\linewidth]{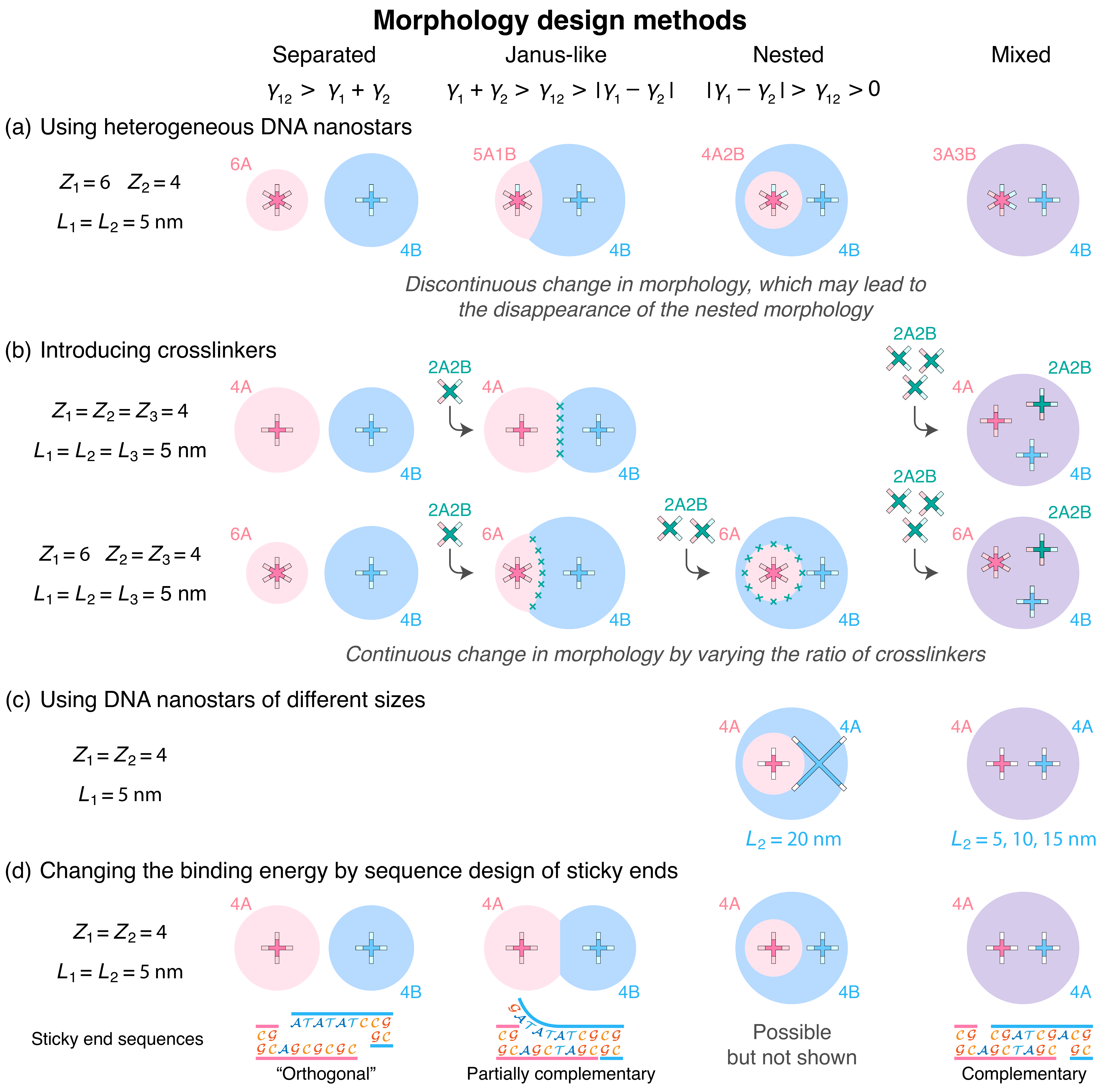}
	\caption{
    Schematics illustrating different strategies for designing morphologies in DNA-nanostar solutions. The DNA nanostar(s) depicted in each dense phase represent the primary constituent component(s).
    (a)~Using heterogeneous DNA nanostars to generate distinct morphologies (see Figs.~\ref{fig4} and~S8 for details).
    (b)~Introducing heterogeneous DNA nanostars as crosslinkers to fuse two initially separate droplets and alter their overall morphology (Figs.~\ref{fig4.5} and~S9).
    (c)~Employing nanostars of different sizes to produce nested morphologies (Fig.~S10).
    (d)~Tuning the binding energy between distinct sticky-end types through sequence design to control the resulting morphologies (Fig.~S11).
    The sticky-end sequences are adapted from Ref.~\cite{jeon2020sequence} for illustration. In the simulations presented here, sequence specificity is not modeled explicitly; instead, the binding energy of coarse-grained sticky ends is adjusted to mimic sequence effects.
    }
	\label{fig5}
\end{figure*}

\subsection{Morphology of two-component systems}

As shown in Fig.~\ref{fig4}, we investigate a broad set of DNA-nanostar combinations and characterize the resulting morphologies using large-scale molecular dynamics simulations ({\it Methods}). The morphologies exhibit a consistent trend as the distribution of sticky-end types is varied, allowing them to be organized into a unified morphology spectrum. Within this spectrum, Janus and nested morphologies occupy intermediate regimes between fully separated condensates and a single mixed dense phase. When the two nanostar types bear entirely distinct sticky ends that do not attract each other, they form condensates that remain separate and do not make contact. When one sticky end on the pink nanostars is replaced by a yellow end identical to those on the blue nanostars, the two condensates come into contact, forming Janus-like structures with a shared interface. For most combinations involving different valences, further replacement of pink sticky ends by yellow ones causes the two condensates to merge directly into a single dense phase. Nested structures appear only within a narrow range of combinations exhibiting large valence differences between the two nanostar species. These trends are consistent with the predictions based on interfacial-energy calculations. The condition for nested morphologies, \(\gamma_1 - \gamma_2 > \gamma_{12}\), requires a substantial contrast between \(\gamma_1\) and \(\gamma_2\) while maintaining a small interfacial energy \(\gamma_{12}\). Figure~\ref{fig2}b shows that a large \(|\gamma_1 - \gamma_2|\) can be achieved by varying nanostar valence or size at a fixed binding energy, whereas Fig.~\ref{fig3} demonstrates that \(\gamma_{12}\) can be minimized by adjusting the sticky-end distribution. For many combinations in Fig.~\ref{fig4}, the interfacial energies \(\gamma_i\) are comparable because the valences of the two nanostar types differ only slightly. Consequently, the interfacial energy \(\gamma_{12}\) must be extremely small to enable the formation of a nested morphology. However, this condition is difficult to achieve, as small changes in sticky-end composition cause discrete jumps in \(\gamma_{12}\), as shown in Fig.~\ref{fig3}, where \(\gamma_{12}\) abruptly disappears as the two condensates merge into a single dense phase. As a result, nested morphologies vanish from the morphology spectrum for most combinations with similar valences.

\subsection{Crosslinking two dense phases}

Although modifying sticky ends is straightforward in simulations, it requires DNA sequence redesign and the synthesis of new nanostars in experiments. Alternatively, crosslinkers offer a more experimentally accessible means of tuning interfacial energies \cite{jeon2020sequence, sato2020sequence}. This approach can also mitigate the absence of nested morphologies observed in some combinations in Fig.~\ref{fig4}, which arises from discrete changes in interfacial energies. As shown in Fig.~\ref{fig4.5}, we introduce a type of crosslinker nanostar that interacts with two homogeneous nanostar species, 6A and 4B. The crosslinker nanostar contains 2 sticky ends~A and 2 sticky ends~B, enabling it to fuse the two otherwise separated droplets. The morphological evolution follows the same trend as in Fig.~\ref{fig4}, showing that varying the crosslinker ratio has an effect similar to altering the sticky-end composition in modulating the interfacial energies. To examine the spatial distribution of crosslinkers, we also perform one-dimensional diffusive phase-field simulations (Fig.~\ref{fig4.5}e–h; details in Appendix~C, {\it Supporting Information}). The results show that crosslinkers (yellow solid lines) preferentially localize at the interface of the pink condensate, acting as surfactants that lower interfacial energy and promote adhesion to the adjacent blue phase. As the crosslinker ratio is a macroscopic thermodynamic variable that can be tuned continuously, it modulates the interfacial energies smoothly. This continuity makes it easier to satisfy the condition for nested morphologies, \(\gamma_1 - \gamma_2 > \gamma_{12}\), prior to complete mixing, provided that \(\gamma_1 - \gamma_2 \neq 0\). Previous experiments \cite{jeon2020sequence, sato2020sequence} reported only Janus-like morphologies when using crosslinkers, because the two components examined were compositionally similar.

\section{Discussion}

In this study, we systematically investigate how constituent biomolecules can be utilized to tune interfacial energies and thereby generate diverse morphologies, focusing on solutions of DNA nanostars.
We combine large-scale coarse-grained molecular dynamics simulations with a generalized Flory–Huggins framework to understand and predict the phase behavior of DNA-nanostar solutions. Figure~\ref{fig5} summarizes several strategies for designing and constructing distinct condensate morphologies using DNA nanostars, including heterogeneous nanostar design (Fig.~\ref{fig5}a), the introduction of crosslinkers (Fig.~\ref{fig5}b), the use of differently sized nanostars (Fig.~\ref{fig5}c), and sticky-end modification (Fig.~\ref{fig5}d). Separated morphologies can be realized by assigning completely distinct sticky ends to each nanostar type, ensuring that the resulting condensates remain independent and noninteracting. For adjacent dense phases that permit dynamic mass exchange—including Janus-like and nested morphologies—moderate attractive interactions between components are required to stabilize the interface while preventing complete mixing. Such interactions can be introduced either by adding identical sticky ends to both nanostar types (Fig.~\ref{fig5}a) or by incorporating crosslinker nanostars that carry sticky ends from both original species (Fig.~\ref{fig5}b). In addition, distinct sticky ends can be designed to attract each other through sequence complementarity (Fig.~\ref{fig5}d), providing a direct means to tune intermolecular interactions \cite{jeon2020sequence}. In principle, various morphologies can be achieved simply by tuning the binding energy between different sticky ends without altering structural parameters such as valence or size. However, the accessible range of binding energies is constrained by the thermodynamics of DNA hybridization and thus cannot be varied arbitrarily. Moreover, combining low valence with strong binding can result in gel-like networks with nearly frozen internal connectivity. These design constraints mirror those encountered in biopolymer condensates, where cells often employ alternative regulatory strategies—such as modulating valence—to control intermolecular interactions and maintain dynamic, liquidlike phases essential for biological function \cite{banani2017biomolecular, soding2020mechanisms, jeon2025emerging}.

To realize the relatively rare nested morphology, one must increase the interfacial-energy contrast, quantified as \(|\gamma_1-\gamma_2|\), while keeping \(\gamma_{12}\) small but finite. The disparity in \(\gamma_i\) can be enhanced by increasing differences in valence (Fig.~\ref{fig5}a,~b), size (Fig.~\ref{fig5}c), binding energy, or other parameters shown in Fig.~\ref{fig2}. However, maintaining a small \(\gamma_{12}\) without triggering complete fusion between the two dense phases remains challenging. The interfacial energy \(\gamma_{12}\) can be tuned by modifying the distribution of sticky-end types (Fig.~\ref{fig5}a), adjusting the ratio of crosslinkers (Fig.~\ref{fig5}b), or directly varying the binding energy between heterotypic sticky-end pairs (Fig.~\ref{fig5}d). Notably, the nested morphology can be readily achieved by simply increasing the size difference between the two homogeneous components—without introducing additional sticky-end types or fine-tuning \(\gamma_{12}\)—as shown in Figs.~\ref{fig5}c and~S10. In this scenario, segregation of smaller nanostars from larger ones reduces the system’s internal energy, promoting relaxation toward a free-energy minimum in which the condensate becomes encapsulated by a phase rich in larger nanostars.

Finally, building on these design principles for three-phase morphologies and the Flory–Huggins framework, one can extend the prediction and design of more complex structures involving four or more coexisting phases of DNA nanostars. In principle, the Flory–Huggins model described by Eqs.~(\ref{Eq:fx}–\ref{Eq:Florychiij}) can be applied to systems with an arbitrary number of components, requiring only the essential parameters for each species (e.g., \(z_{i,\rm eff}, M_{i,\rm eff}\)). The intercomponent interaction parameters \(\chi_{ij}\) are then automatically determined from Eq.~(\ref{Eq:Florychiij}), demonstrating the versatility of this Flory–Huggins framework. Following the graph-theoretical approach to morphology design \cite{mao2020designing}, each morphology can be mapped to a specific topological relationship among the phases and represented by a {\it graph}. Each complex graph can be decomposed into subgraphs representing triplets of coexisting phases, with each triplet corresponding to a specific interfacial-energy inequality \cite{mao2020designing}. Using the methodology developed in this work, each subgraph can be materialized through the rational design of DNA nanostar components. The desired higher-order morphology can then be realized by integrating these triplet elements. By combining the theoretical and computational framework established here with the graph-theoretical design approach, more intricate morphologies with well-orchestrated functions can ultimately be realized experimentally.

\section{Methods}
\subsection{Development of the coarse-grained model}

In the coarse-grained model (Fig.~\ref{fig1}e–g), both bonded and non-bonded interactions are specified. The bonded interactions account for the energies associated with {\it bonds} and {\it angles}, as defined in LAMMPS \cite{plimpton1995fast}, corresponding to linear springs for chain stretching and angular springs for chain bending. Both types of springs are described by harmonic potentials, i.e., \(E_{\rm Bond} = \frac{1}{2} k_d (d - d_0)^2\) and \(E_{\rm Angle} = \frac{1}{2} k_\theta (\theta - \theta_0)^2\), where \(d\) is the bond length, \(d_0\) is the equilibrium bond length, \(\theta\) is the bond angle, \(\theta_0\) is its equilibrium value, and \(k_d\) and \(k_\theta\) are the corresponding spring constants. From a continuum perspective \cite{liu2022design, liu2021multilayer, liu2022thedesign}, the spring constants are given by \(k_d = \frac{EA}{d_0}\) and \(k_\theta = \frac{EI}{d_0}\), where \(EA\) and \(EI\) denote the tensile and bending stiffness of the chain, respectively.

We choose the equilibrium bond length \(d_0 \equiv 1~{\rm nm}\), which is much smaller than the persistence length of double-stranded DNA [\(\ell_{\rm p} \approx 50~{\rm nm}\) \cite{geggier2010sequence}]. The mass of each bead is set to \(m = 2000~{\rm amu}\). Given that the tensile stiffness of double-stranded DNA is \(EA \approx 500k_{\rm B}T~{\rm nm^{-1}}\) \cite{marko1995stretching}, we assign \(k_d = 500k_{\rm B}T~{\rm nm^{-2}}\) for all bonds. Three types of bond angles are defined in the model:  (1)~angles along DNA arms (black in Fig.~\ref{fig1}e);  (2)~angles along sticky ends (blue); and  (3)~angles at central nodes (red).  For DNA arms, the angular spring constant is set to \(k_\theta^{(1)} = 50k_{\rm B}T\), consistent with the bending rigidity of double-stranded DNA, \(EI = \ell_{\rm p} k_{\rm B}T\) \cite{geggier2010sequence}. Angles on sticky ends are much softer than those on the arms, since the sticky ends are composed of single-stranded DNA and include an additional unpaired base, which facilitates reversible unbinding and ensures droplet fluidity \cite{jeon2020sequence, nguyen2017tuning}. Accordingly, we set \(k_\theta^{(2)} \equiv k_{\rm B}T\). For type~3 angles, there are \(Z\) such angles around each central node (where \(Z\) is the number of arms), as only adjacent DNA arms are considered to form interacting angles. We assign \(k_\theta^{(3)} \equiv 50k_{\rm B}T\), assuming that the base pairs at the node are fully paired and thus relatively rigid [see Ref.~\cite{nguyen2017tuning} for details]. The node, however, remains flexible in other directions involving non-adjacent arms. For instance, a 4-arm nanostar can fold its opposite arms without changing any of the angles defined around the node. The equilibrium angles for types~1–3 are \(\theta_0^{(1)} = \theta_0^{(2)} = 180^\circ\) and \(\theta_0^{(3)} = 180^\circ/Z\), indicating that the planar configuration of DNA nanostars is taken as the zero-energy state. Note that in Fig.~\ref{fig2}c,~d, only the bending rigidity of DNA arms, \(k_\theta^{(1)}\), is varied to examine its effect on interfacial energy.

The non-bonded interactions include both Coulombic and sticky-end interactions. Since DNA nanostars exist in an ionic solution, we use the Yukawa potential to describe the screened Coulomb interactions between {\it hollow black beads} as illustrated in Fig.~\ref{fig1}. The Yukawa potential is expressed as \(E_{\rm Elec} = \frac{q^2}{4\pi \epsilon_0 \epsilon_r r} e^{-\kappa r}\), where \(q = -6e\) (\(e = 1.6\times10^{-19}~{\rm C}\)) is the charge carried by each black bead (since it represents approximately 3~bp of DNA and each base pair contributes two electrons), \(\epsilon_0 = 8.854\times10^{-12}~{\rm F\,m^{-1}}\) is the vacuum permittivity, \(\epsilon_r = 80\) is the relative permittivity of water, \(r\) is the distance between two black beads, and \(\lambda_{\rm D} \equiv \kappa^{-1} = \sqrt{\frac{\epsilon_0 \epsilon_r k_{\rm B}T}{2e^2 I_{\rm ion}}}\) is the Debye screening length for a monovalent electrolyte, where \(I_{\rm ion}\) denotes the effective ionic concentration (ionic strength) of the solution \cite{russel1991colloidal}. The Debye length is typically on the order of 1~nm for \(I_{\rm ion} \sim 100~{\rm mM}\) in physiological conditions \cite{liu2017ionic}. Therefore, we set \(\lambda_{\rm D} = 2~{\rm nm}\) for all simulations (except those in Fig.~\ref{fig2}c,~d) and apply a cutoff of 3~nm for the Yukawa potential.

The interactions between sticky ends depend on whether the ends are identical or distinct. Identical sticky ends attract each other through complementary base pairing, modeled as attractions between differently colored beads, while beads of the same color repel each other (Fig.~\ref{fig1}). These interactions ensure that only two complementary sticky ends can bind simultaneously. Both attractive and repulsive interactions are represented by soft-core potentials that permit bead overlap, expressed as \(E_{\rm Att} = -\frac{E_0}{2}\left(1 + \cos\frac{\pi r}{r_{\rm cut}}\right)\) and \(E_{\rm Rep} = \frac{E_1}{2}\left(1 + \cos\frac{\pi r}{r_{\rm cut}}\right)\), both valid for \(r < r_{\rm cut}\) and set to zero otherwise. Here, \(E_0 = \frac{1}{2}|\Delta E|\) is determined by the sticky-end binding energy \(\Delta E\), \(E_1 = 50k_{\rm B}T\) represents the energy penalty for overlapping beads of the same color, and \(r_{\rm cut} = 1~{\rm nm}\) is the cutoff distance for both potentials. Between distinct (non-complementary) sticky ends, only repulsive interactions (\(E_{\rm Rep}\)) are defined, irrespective of bead color, reflecting the absence of hybridization and the dominant steric repulsion. For simplicity, the {\it colored beads} on sticky ends also interact repulsively with the {\it hollow black beads} on DNA arms via the same potential, \(E_{\rm Rep} = \frac{E_1}{2}\left(1 + \cos\frac{\pi r}{r_{\rm cut}}\right)\).

\subsection{Simulation procedure}

A Langevin thermostat is employed to regulate the temperature and dynamics of the simulations. The timestep is set to \(\Delta t = 0.426~{\rm ps}\), which is sufficiently small to resolve all vibrational frequencies associated with bonded interactions. The velocity relaxation time of the Langevin dynamics, defined as \(\tau = \frac{m}{\zeta}\) (where \(\zeta\) is the damping coefficient), is chosen to be \(\tau = 10^3\Delta t\). The temperature is maintained at \(T = 298~{\rm K}\) for all simulations, while the sticky-end binding energy \(\Delta E\) is varied to mimic the effect of temperature changes. Periodic boundary conditions are applied in all three spatial directions for every simulation.

For calculating the density and interfacial energy of single-component DNA nanostar systems (Fig.~\ref{fig2}), the simulation box dimensions are set to \(600\times100\times100~{\rm nm^3}\) for nanostars with arm length \(L = 5~{\rm nm}\) and \(1200\times200\times200~{\rm nm^3}\) for \(L = 10~{\rm nm}\). Because larger nanostars produce broader interfaces, the system size is increased accordingly to reduce finite-size effects. Each simulation contains 2500 DNA nanostars, initially placed randomly in the central region of the box, and equilibrated for \(10^7\)~steps. For each parameter set, five independent samples with distinct initial configurations are equilibrated and then simulated for approximately \(10^8\)~steps to compute the interfacial energy using the Kirkwood–Buff formula \cite{kirkwood1949statistical}: \(\gamma = \frac{L_x}{2} \left\langle P_x - \frac{P_y + P_z}{2} \right\rangle\), where \(L_x = 600~{\rm nm}\) is the box length along the \(x\)-direction, \(P_i\) denotes the diagonal components of the pressure tensor, and the factor of \(1/2\) accounts for the two interfaces present in the simulation box.

For two-component systems, the interfacial energy is computed using a similar procedure. The simulation box size is expanded to \(1000\times100\times100~{\rm nm^3}\) to accommodate 4000 nanostars of type~1 on the left side and 4000 nanostars of type~2 on the right side (Fig.~S7). Interfacial energies are obtained through regional averaging of the pressure difference in the Kirkwood–Buff formula, expressed as \(\gamma^{(i)} = L_x^{(i)} \left\langle P_x - \frac{P_y + P_z}{2} \right\rangle_{\text{Region}}^{(i)}\), where \(\gamma^{(i)}\) denotes the interfacial energy of the interface within region~\(i\), \(L_x^{(i)}\) is the length of that region along the \(x\)-direction, and \(\langle \dots \rangle_{\text{Region}}^{(i)}\) represents the spatial average of the pressure tensor components over that region. Five independent systems are equilibrated for \(10^7\)~steps and subsequently run for more than \(2.5\times10^8\)~steps to obtain time-averaged results. To confirm the morphological outcomes, large-scale simulations shown in Figs.~\ref{fig4},~\ref{fig4.5},~and~\ref{fig5} are performed for up to \(6\times10^8\)~steps until the configurations reach steady states. Detailed parameters for these simulations are provided in Figs.~S8–S11 and Table~S2.

\section*{Acknowledgement}
This research was primarily supported by NSF through Princeton University’s Materials Research Science and Engineering Center (DMR-2011750). We would like to acknowledge useful discussions with Omar Saleh.

\bibliography{nanostar_ref}

\end{document}


\preprint{APS/123-QED}

\title{Supporting Information: Predicting the Interfacial Energy and Morphology of\\DNA Condensates}

\author{Sihan Liu}
 \affiliation{Department of Mechanical and Aerospace Engineering, Princeton University, Princeton, NJ 08544, USA}

\author{Andrej Ko\v{s}mrlj}
 \email{andrej@princeton.edu}
 \affiliation{Department of Mechanical and Aerospace Engineering, Princeton University, Princeton, NJ 08544, USA}
 \affiliation{Princeton Materials Institute, Princeton University, Princeton, NJ 08544, USA}

\date{\today}

\maketitle

{\bf  This file includes:}

\quad Appendix~\ref{sec1}. Flory–Huggins theory for DNA nanostar solutions.

\quad Appendix~\ref{sec3}. Derivation of the interfacial energy in a multi-component system.

\quad Appendix~\ref{sec2}. Details of diffusive phase-field simulations.

\quad Fig.~\ref{lattice}. Schematics of the generalized lattice model for DNA-nanostar solutions.

\quad Fig.~\ref{figs1.5}. Statistics of paired sticky ends.

\quad Fig.~\ref{sequence}. Interfacial energy of heterogeneous nanostars with different sticky-end arrangements.

\quad Fig.~\ref{figs phase diagram}. Comparison of the interfacial energies shown in Fig.~4c with the phase-field simulation results.

\quad Fig.~\ref{Meff}. Interfacial energy as a function of the Flory interaction parameter \(\chi\) in a binary system.

\quad Fig.~\ref{figs1}. Power-law scaling of the interfacial energy near the critical point.

\quad Fig.~\ref{figs2}. Simulation snapshots of two-component systems to calculate interfacial energy.

\quad Fig.~\ref{figs3}. Designing different morphologies by using heterogeneous DNA nanostars.

\quad Fig.~\ref{figs4}. Designing different morphologies by introducing crosslinkers.

\quad Fig.~\ref{figs5}. Designing different morphologies by using differently sized DNA nanostars.

\quad Fig.~\ref{figs6}. Designing different morphologies by changing the binding energy of sticky ends.

\quad Table~\ref{table:FH}. Summary of the fitting parameters in the Flory–Huggins theory.

\quad Table~\ref{table:md}. Summary of the molecular dynamics simulation parameters.

\section{Flory–Huggins theory for DNA nanostar solutions} \label{sec1}

To quantitatively elucidate how valence, arm length, bending rigidity, and Debye length influence the phase behavior shown in Fig.~2, we employ the Flory–Huggins theory \cite{flory1942thermodynamics, huggins1941solutions} for DNA-nanostar solutions. As illustrated in Fig.~\ref{lattice}, we introduce a generalized lattice model in which each DNA nanostar is represented as an effective particle occupying a single lattice site. Each site may either host a DNA nanostar or remain vacant to represent solvent or void space, while neighboring nanostars can form a bond or remain unbound. The center of each nanostar is located at the center of its lattice cell, with its DNA arms fluctuating around the central node. Denoting the total number of lattice sites as \(N\) and the number of nanostar molecules as \(N_{\rm star}\), the total free energy of mixing, \(F\), can be written as \cite{flory1942thermodynamics, huggins1941solutions}
\begin{align} \label{Eq:F}
    \frac{F}{Nk_{\rm B}T} =
    \frac{\phi}{M_{\rm eff}} \ln \phi
    + \left(1-\phi\right) \ln (1-\phi)
    +\chi \phi (1-\phi),
\end{align}
where \(k_{\rm B}\) is the Boltzmann constant, \(T\) is the temperature, \(\phi \equiv N_{\rm star}/N\) is the volume fraction of lattice sites occupied by DNA nanostars, and \(\chi\) is the Flory interaction parameter. The entropy of mixing in Eq.~(\ref{Eq:F}) includes a correction factor \(M_{\rm eff}\), which governs the asymmetry of the binodal lines in Fig.~2 and is analogous to the polymerization factor in polymer solutions \cite{flory1942thermodynamics, huggins1941solutions}. This correction factor arises because DNA nanostars are not perfectly isotropic point particles and cannot interact with all their neighbors. Unlike the regular solution model, which assumes a single configuration in the reference demixed state, a nanostar network prior to mixing exhibits substantial internal conformational entropy, as shown in Fig.~\ref{lattice}a, similar to linear polymer solutions \cite{flory1942thermodynamics, huggins1941solutions}. In general, nanostars with higher valence and smaller size behave more like isotropic point particles that can interact with all neighbors, corresponding to \(M_{\rm eff} = 1\). From fitting the Flory–Huggins binodal lines to the simulation results in Fig.~2a, we find \(M_{\rm eff}\) to be approximately 4.44, 5.94, and 7.53 for \(L = 5\)~nm with \(Z = 6, 4,\) and~3, respectively, and 9.33 for \(L = 10\)~nm with \(Z = 4\).

Since each nanostar occupies a single lattice site, the lattice spacing \(a\) characterizes the effective molecular size of the nanostars. Accordingly, the overall molecular number density is expressed as \(\rho = N_{\rm star}/V = \phi \rho_0\), where \(V\) is the total system volume and \(\rho_0 = N/V \sim a^{-3}\) represents the number density of lattice sites. Substituting \(\rho\) into Eq.~(\ref{Eq:F}), the free-energy density, \(f \equiv F/V\), can be written as
\begin{align} \label{Eq:f}
    \frac{f}{k_{\rm B}T} =
    \frac{\rho}{M_{\rm eff}} \ln \frac{\rho}{\rho_0}
    + \left(\rho_0-\rho\right) \ln \left(1-\frac{\rho}{\rho_0}\right)
    +\chi \rho \left(1-\frac{\rho}{\rho_0} \right).
\end{align}
Equation~(\ref{Eq:f}) shows that a smaller \(\rho_0\) narrows the width of the coexistence region when plotted as a function of the molecular number density \(\rho\) (see Fig.~2). We measure \(\rho_0\) to be approximately \(2.02\times10^{-3}\ {\rm nm^{-3}}\), \(1.91\times10^{-3}\ {\rm nm^{-3}}\), and \(1.52\times10^{-3}\ {\rm nm^{-3}}\) for \(L = 5\ {\rm nm}\) with \(Z = 6, 4,\) and~3, respectively, and \(0.35\times10^{-3}\ {\rm nm^{-3}}\) for \(L = 10\ {\rm nm}\) with \(Z = 4\). These results demonstrate that \(\rho_0 \sim a^{-3}\) decreases with increasing nanostar size and increases with valence due to enhanced network connectivity.
A reasonable approximation for \(a\) can be derived from the mean-square end-to-end distance of the DNA arms, \(a \approx 2\langle R^2\rangle^{1/2}\), where \(\vec{R}\) is the end-to-end vector of a single DNA arm in a given configuration, and the brackets denote the ensemble average. For worm-like chains, the mean-square end-to-end distance is given by \cite{rubinstein2003polymer}
\begin{align} \label{Eq:R2}
    \langle R^2\rangle &= 2\ell_{\rm p}L - 2\ell_{\rm p}^2
    \left(
    1-e^{-L/\ell_{\rm p}}
    \right)
    \sim
    \left\{
    \begin{array}{ll}
    \displaystyle
    L^2,
    &  L\ll\ell_{\rm p}, \\
    \displaystyle
    2\ell_{\rm p}L,
    &  L\gg\ell_{\rm p}.
    \end{array}
    \right.
\end{align}
Here, \(\ell_{\rm p}\) denotes the persistence length of the polymer chain, which is approximately 50~nm for double-stranded DNA \cite{geggier2010sequence}. Since the DNA nanostars considered in this work and related experiments are relatively small (\(L \le 10~{\rm nm}\)) \cite{jeon2020sequence, sato2020sequence}, we focus on the regime \(L \ll \ell_{\rm p}\). Although the Debye screening length can influence the effective persistence length of a charged polymer \cite{chen2012ionic}, this effect is negligible in the present case, where the DNA arms remain nearly straight.

\renewcommand{\thefigure}{S\arabic{figure}}
\begin{figure*}[!t]
	\centering
	\includegraphics[width=0.95\linewidth]{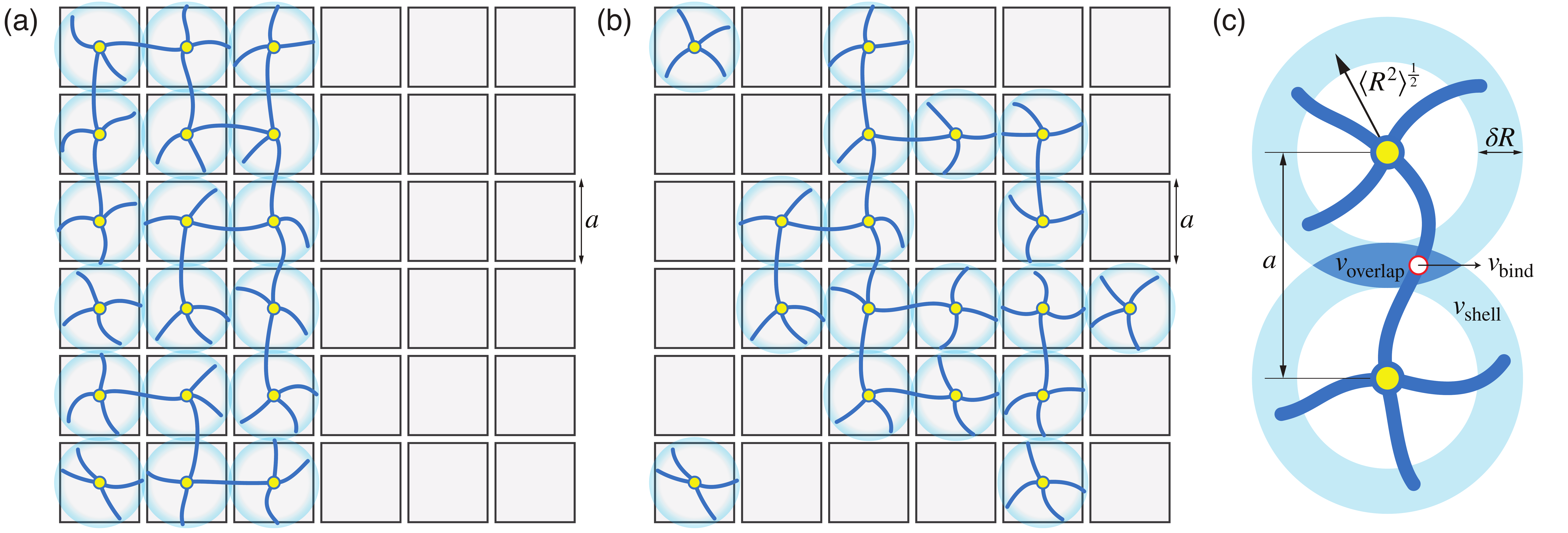}
	\caption{
    (a,~b) Schematics of the generalized lattice model for DNA-nanostar solutions. Panel (a) shows the reference (demixed) state, which includes the conformational entropy of individual nanostars, while panel (b) illustrates the mixed state. Each nanostar occupies a single lattice site and interacts with its nearest neighbors. Empty sites are filled with solvent, while each occupied site also contains solvent surrounding the nanostar. The lattice spacing~\(a\) represents the effective molecular size of a DNA nanostar. The central junction (yellow node) is located at the center of the site, and its sticky ends are distributed within a blue spherical shell that partially overlaps with neighboring shells.  
    (c)~Schematic of the binding between two adjacent DNA nanostars. The sticky ends are distributed within a spherical shell of volume \(v_{\rm shell}\), and binding occurs only within the overlap region shared by the two shells, with a volume \(v_{\rm overlap}\). Each sticky end occupies a characteristic binding volume \(v_{\rm bind}\), within which it can interact with and hybridize with another complementary sticky end.
    }
	\label{lattice}
\end{figure*}

In Eqs.~(\ref{Eq:F},~\ref{Eq:f}), the Flory interaction parameter \(\chi\) determines whether phase separation occurs. When \(\chi\) is large, the free-energy density \(f\) becomes unstable over an intermediate range of molecular number densities \(\rho\), driving the system to separate into a dilute phase and a dense phase whose respective densities are defined by the coexistence line (Fig.~2a). In the corresponding lattice model, the Flory interaction parameter \(\chi\) is defined as \cite{shell2015thermodynamics}
\begin{align} \label{Eq:chi_def}
    \chi &= \frac{2(z_{\rm eff}-2)w_{10}-(z_{\rm eff}-2)w_{11}-z_{\rm eff}w_{00}}{2k_{\rm B}T}.
\end{align}
Here, \(z_{\rm eff}\) denotes the effective coordination number of the generalized lattice, which reflects the packing number of DNA nanostars and increases with their valence \(Z\). Superscripts “1” and “0” correspond to lattice sites occupied by nanostars and unoccupied sites filled with solvent, respectively, while \(w_{11}\), \(w_{00}\), and \(w_{10}\) represent the effective interaction energies between neighboring sites of the respective types. Note that \(w_{11}\) is not equal to the sticky-end binding energy \(\Delta E\), since two nanostars do not necessarily bind and the conformational entropy between nanostars contributes. Although DNA nanostars differ intrinsically from linear polymers, the asymmetric coexistence line in Fig.~2 suggests that they can effectively polymerize into dynamic clusters with an average polymerization factor \(M_{\rm eff} > 1\). Consequently, we subtract 2 from \(z_{\rm eff}\) in the definition of \(\chi\) for terms involving occupied sites to maintain consistency with the conventional Flory–Huggins framework for polymer solutions, where interactions between a monomer and its two covalently bonded neighbors along the same chain are excluded \cite{shell2015thermodynamics}. This modification does not alter subsequent analyses, as \(z_{\rm eff}-2\) can still be treated as a single fitting parameter, but it provides a more physically reasonable value for \(z_{\rm eff}\). Because the occupied sites also include solvent surrounding the DNA nanostars, solvent–solvent interactions cancel out in Eq.~(\ref{Eq:chi_def}) when taking the difference between the mixed configuration (Fig.~\ref{lattice}b) and the reference demixed configuration (Fig.~\ref{lattice}a). This cancellation reflects the fact that the Flory–Huggins free energy accounts only for the relative change in energy upon mixing. Consequently, only the energy terms associated with DNA nanostars contribute to Eq.~(\ref{Eq:chi_def}). These contributions can be estimated from the partition functions of pairs of neighboring lattice sites,
\renewcommand{\thefigure}{S\arabic{figure}}
\begin{figure*}[b!]
	\centering
	\includegraphics[width=0.9\linewidth]{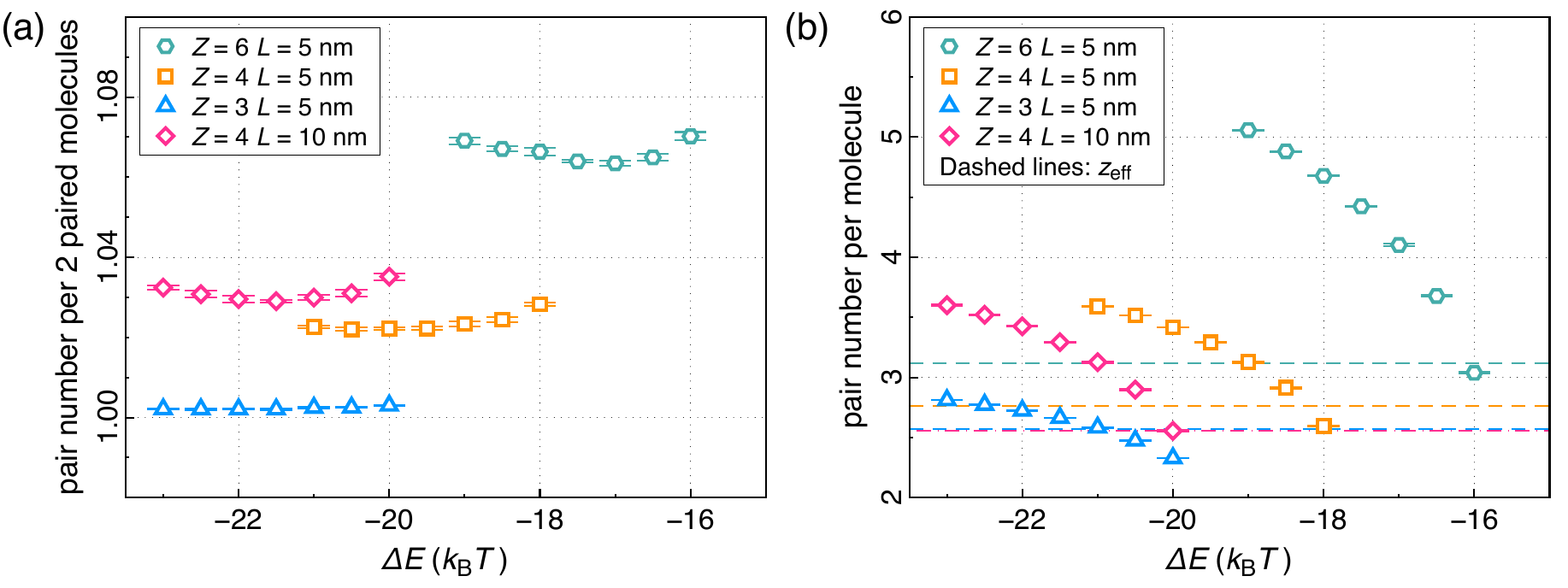}
	\caption{Statistics of paired sticky ends. (a) The average number of sticky-end pairs formed between two paired DNA nanostars, which must exceed 1 but remains very close to unity. This supports our approximation in the Flory–Huggins model that only a single sticky-end pair forms between two neighboring nanostars. (b) The average number of paired sticky ends per DNA nanostar. The dashed line indicates the coordination number \(z_{\rm eff}\) in the generalized lattice model for different structures, obtained by fitting the phase diagram (Fig. 2a) using the Flory–Huggins theory. Notably, \(z_{\rm eff}\) approaches the number of paired sticky ends per nanostar only near the critical point. Error bars represent the standard error of the mean (SEM) computed from five independent simulations.}
	\label{figs1.5}
\end{figure*}
\begin{subequations} \label{Eq:Z}
\begin{align} 
    \mathcal{Z}_{11} &\approx \Omega_{\rm unbind} + \Omega_{\rm bind} e^{-\Delta E/(k_{\rm B}T)}, \label{Eq:Z11}
    \\
    \mathcal{Z}_{10} &= \Omega_{\rm star},
    \\
    \mathcal{Z}_{00} &= 1,
\end{align}
\end{subequations}
where \(\Omega_{\rm unbind}\) and \(\Omega_{\rm bind}\) denote the total numbers of configurations of two neighboring nanostars in the unbound and bound states, respectively; \(\Delta E\) is the binding energy of a sticky-end pair; and \(\Omega_{\rm star}\) denotes the total number of configurations of a single nanostar. For simplicity, we consider the case in which only a single pair of DNA arms can bind between two adjacent nanostars. Self-binding is excluded because the arms are relatively rigid and experience electrostatic repulsion along the charged DNA backbone, despite the attractive interactions between complementary sticky ends. This approximation is supported by the binding statistics in Fig.~\ref{figs1.5}, which show that even for two bound 6-arm nanostars, only 1.07 sticky-end pairs form on average. The probability of self-binding is extremely low in our simulations: the number of self-bound pairs per nanostar is below \(\sim10^{-5}\) when \(L = 10~\mathrm{nm}\) and exactly zero when \(L = 5~\mathrm{nm}\). Based on Eqs.~(\ref{Eq:Z}), the effective free energies between two neighboring lattice sites are given by
\begin{subequations}
\begin{align} \label{Eq:w}
    w_{11} &= -k_{\rm B}T\ln\mathcal{Z}_{11} \approx -k_{\rm B}T 
    \ln\left[
    \Omega_{\rm unbind} + \Omega_{\rm bind} e^{-\Delta E/(k_{\rm B}T)}
    \right],
    \\
    w_{10} &= -k_{\rm B}T\ln\mathcal{Z}_{10} = -k_{\rm B}T 
    \ln \Omega_{\rm star},
    \\
    w_{00} &= -k_{\rm B}T\ln\mathcal{Z}_{00} = 0
    .
\end{align}    
\end{subequations}
In the following calculations of the conformational entropy, we assume that the centers of two nanostars are separated by the lattice spacing \(a\) (Fig.~\ref{lattice}c), consistent with the lattice model. Since the number of unbound configurations is generally much larger than that of bound ones, we approximate it by the total number of configurations accessible to two nanostars separated by a constant distance~\(a\), given by
\begin{align}
    \Omega_{\rm unbind} = \Omega_{\rm star}^2 - \Omega_{\rm bind} \approx \Omega_{\rm star}^2.
\end{align}
Therefore, the Flory interaction parameter in Eq.~(\ref{Eq:chi_def}) reduces to
\begin{align} \label{Eq:chi_simplified}
    \chi &= -\frac{(z_{\rm eff}-2)(w_{11}-2w_{10})}{2k_{\rm B}T}
    \approx \frac{z_{\rm eff}-2}{2} 
    \ln\left[
    1 + e^{-(\Delta E - T\Delta S)/(k_{\rm B}T)}
    \right],
\end{align}
where
\begin{align}
    \Delta S = k_{\rm B} \ln \left(\frac{\Omega_{\rm bind}}{\Omega_{\rm unbind}}\right)
\end{align}
represents the loss of conformational entropy upon binding, arising from the spatial constraints imposed on the paired arms.
According to the partition function \(\mathcal{Z}_{11}\) in Eq.~(\ref{Eq:Z11}), the probability of forming a single sticky-end pair between two nanostars is given by
\begin{align}
    p_{\rm bind} &\approx \frac{\Omega_{\rm bind}e^{-\Delta E/(k_{\rm B}T)}}{\Omega_{\rm unbind}+\Omega_{\rm bind}e^{-\Delta E/(k_{\rm B}T)}}
    =
    \frac{e^{-(\Delta E - T\Delta S)/(k_{\rm B}T)}}{1+e^{-(\Delta E - T\Delta S)/(k_{\rm B}T)}}
    .
\end{align}
For DNA nanostars, we find that \(p_{\rm bind}\approx1\) once phase separation occurs at large \(|\Delta E|\), with \(p_{\rm bind,c}\gtrsim0.9\) already at the critical point \(\Delta E_{\rm c}\). The value of \(p_{\rm bind,c}\) can later be estimated by fitting the binodal line to extract \(\Delta E_{\rm c} - T\Delta S\), which in turn determines \(p_{\rm bind,c}\). Therefore, in the phase-separated regime, we take \(\chi \approx - (z_{\rm eff}-2) (\Delta E - T\Delta S)/(2k_{\rm B}T)\) in Eq.~(\ref{Eq:chi_simplified}), as the bound-state term dominates the partition function. To compute the conformational entropy change upon binding, we explicitly evaluate \(\Omega_{\rm unbind}\) and \(\Omega_{\rm bind}\). The number of configurations of the unbound state can be decomposed as
\begin{align}
    \Omega_{\rm unbind}  \approx \Omega_{\rm star}^2 = \Omega_{\rm arm}^{2}\Omega_{\rm rest}^2,
\end{align}
where \(\Omega_{\rm star} = \Omega_{\rm arm}\Omega_{\rm rest}\) denotes the total number of configurations of a single nanostar, \(\Omega_{\rm arm}\) is the number of accessible configurations of one DNA arm, and \(\Omega_{\rm rest}\) accounts for the configurational degrees of freedom of the remaining arms, conditioned on the position of that arm. Note that \(\Omega_{\rm rest}\) inherently includes steric repulsion among DNA arms.
On the other hand, the total number of bound states can be approximated as
\begin{align} \label{Eq:Omega_bind}
    \Omega_{\rm bind}
    &= Z^2  \Omega_{\rm pair},
    \\
    \label{Eq:Omega_pair}
    \Omega_{\rm pair}
    &\approx
    \left(\Omega_{\rm arm} \cdot \frac{v_{\rm overlap}}{v_{\rm shell}}\right)
    \cdot 
    \left(
    \Omega_{\rm arm} \cdot \frac{v_{\rm bind}}{v_{\rm shell}}
    \right)
    \cdot
    \Omega_{\rm rest}^2
    \notag\\
    &= \frac{ v_{\rm overlap}v_{\rm bind} }{v_{\rm shell}^2}
    \cdot \Omega_{\rm arm}^2\Omega_{\rm rest}^2.
\end{align}
In Eq.~(\ref{Eq:Omega_bind}), the factor \(Z^2\) accounts for the number of possible ways to select one arm from each of the two nanostars to form a pair, and \(\Omega_{\rm pair}\) denotes the number of remaining configurations once a specific pair of DNA arms is selected. A key assumption in estimating \(\Omega_{\rm pair}\) in Eq.~(\ref{Eq:Omega_pair}) is that the endpoint of each DNA arm, denoted by \(\vec R\), is distributed within a spherical shell of volume \(v_{\rm shell}\) and effective thickness \(\delta R \sim (\langle R^2\rangle - \langle |\vec R|\rangle^2)^{1/2}\) (Fig.~\ref{lattice}c). In the rigid-arm limit (\(L \ll \ell_{\rm p}\)), this shell becomes infinitesimally thin, and the probability density of each sticky-end position is taken to be uniform within the shell, \(p(\vec R) = 1/v_{\rm shell}\), and zero outside. Consequently, the configuration number of binding is determined by several characteristic volumes. Since two neighboring nanostars each possess such an endpoint-distribution shell, binding can occur only within their overlap region of volume \(v_{\rm overlap}\) (Fig.~\ref{lattice}c). The number of states available to one DNA arm within the overlap region is thus \(\Omega_{\rm arm}v_{\rm overlap}/v_{\rm shell}\). When another DNA arm from the adjacent nanostar binds to it, this incoming arm further reduces its available states to \(\Omega_{\rm arm}v_{\rm bind}/v_{\rm shell}\), where \(v_{\rm bind}\) denotes the effective binding volume of the sticky end (Fig.~\ref{lattice}c), within which two sticky ends are sufficiently close to hybridize. The final factor \(\Omega_{\rm rest}^2\) in Eq.~(\ref{Eq:Omega_bind}) represents the configurational states of the remaining arms, which are assumed not to interact with the other nanostar, as the overlap region is small and only the two bound arms are considered within it. Combining all contributions, the conformational entropy change upon binding is given by
\begin{align} \label{Eq:S}
    \Delta S = k_{\rm B} \ln \left(\frac{\Omega_{\rm bind}}{\Omega_{\rm unbind}}\right)
    \approx
    k_{\rm B}\ln \left(\frac{Z^2v_{\rm overlap} v_{\rm bind}}{v_{\rm shell}^2}\right),
\end{align}
which is governed by the valence and the characteristic volumes, while the configurational number of each DNA nanostar, \(\Omega_{\rm star}=\Omega_{\rm arm}\Omega_{\rm rest}\), cancels out under the assumption of a homogeneous probability density. Because the shell volume \(v_{\rm shell}\) is much larger than both \(v_{\rm overlap}\) and \(v_{\rm bind}\), the conformational entropy decreases substantially upon binding, although a higher valence \(Z\) mitigates this loss. The effective detection volume of sticky ends, \(v_{\rm bind}\), is assumed to be identical for different sticky-end types and to depend on the Debye screening length. In particular, a shorter \(\lambda_{\rm D}\) leads to a larger \(v_{\rm bind}\), as reduced electrostatic screening facilitates encounters between sticky ends and thereby increases the effective detection volume. The effective shell volume \(v_{\rm shell}\) and the overlap volume \(v_{\rm overlap}\) can be obtained geometrically. For simplicity, we consider the specific configuration shown in Fig.~\ref{lattice}c, in which the outer surface of one shell is tangent to the inner surface of the other. In this geometry, both \(v_{\rm shell}\) and \(v_{\rm overlap}\) depend on the shell radius \(R\) and thickness \(\delta R\). The overlap region in three dimensions consists of two spherical caps (see Fig.~\ref{lattice}c), and the ratio \(v_{\rm overlap}/v_{\rm shell}^2\) can therefore be expressed as
\begin{align} \label{Eq:volume}
    \frac{v_{\rm overlap}}{v_{\rm shell}^2} \approx \frac{
    \frac{\pi }{2}R\delta R ^2
    }{(4\pi R^2 \delta R)^2}
    = \frac{1}{32\pi R^3}
    \approx \frac{1}{32\pi L^3},
\end{align}
which is independent of the shell thickness \(\delta R\). Therefore, in the limit \(\langle R^2\rangle \to L^2\) for \(L \ll \ell_{\rm p}\), the Flory interaction parameter can be approximated as
\begin{align} \label{Eq:chi_DeltaE}
    \chi
    &\approx \frac{z_{\rm eff}-2}{2}
    \left[
    -\frac{\Delta E}{k_{\rm B}T}
    +2\ln Z- 3 \ln L + \ln v_{\rm bind}- \ln (32\pi)
    \right]
    \notag \\
    & \equiv 
    \frac{z_{\rm eff}-2}{2}
    \left[
    -\frac{\Delta E}{k_{\rm B}T}
    +2\ln Z- 3 \ln\overline L + C (v_{\rm bind})
    \right],
\end{align}
where \(\overline L \equiv L/(1~{\rm nm})\) denotes the dimensionless arm length, and \(C\) is a function that incorporates \(v_{\rm bind}\) and therefore depends on the Debye length \(\lambda_{\rm D}\). By fitting the Flory–Huggins binodal lines to the phase diagrams in Fig.~2a, we obtain \(C = -12.6 \pm 0.5\) for all systems with persistence length \(\ell_{\rm p}=50~{\rm nm}\) and Debye length \(\lambda_{\rm D}=2~{\rm nm}\). The effective coordination numbers are found to be \(z_{\rm eff}\approx 3.12\), 2.76, and 2.57 for \(L=5\)~nm with \(Z=6\), 4, and 3, respectively, and \(z_{\rm eff}\approx 2.56\) for \(L=10\)~nm with \(Z=4\). These values are consistent with the expectation that \(z_{\rm eff}\) increases with \(Z\) while remaining smaller than \(Z\). All measured parameters in this section are summarized in Table~\ref{table:FH}.

Equation~(\ref{Eq:chi_DeltaE}) quantitatively captures how the binding energy, temperature, valence, and arm length together regulate the phase behavior of DNA condensates. Because the critical interaction parameter \(\chi_{\rm c}\) is nearly constant, with only weak dependence on \(M_{\rm eff}>4\) (see Fig.~\ref{Meff}), increasing the valence \(Z\) or decreasing the arm length \(L\) lowers the magnitude of the critical binding energy \(\Delta E_{\rm c}\) (i.e., makes \(\Delta E_{\rm c}\) less negative), or equivalently raises the critical temperature required for phase separation—consistent with the coexistence lines in Fig.~2. Equation~(\ref{Eq:chi_DeltaE}) also predicts that \(\chi\) is independent of the bending rigidity of the DNA arms in the rigid limit \(L \ll \ell_{\rm p}\), since \(\delta R(\ell_{\rm p})\) cancels in Eq.~(\ref{Eq:volume}). In contrast, increasing the Debye length reduces the effective binding volume \(v_{\rm bind}\), which decreases the parameter \(C = \ln v_{\rm bind} - {\rm const.}\) and therefore requires a larger \(|\Delta E_{\rm c}|\) to drive condensation.

Note that in Fig.~2c,~d, the coexistence curve and the interfacial energy remain nearly unchanged when the persistence length \(\ell_{\rm p}\) is reduced from 50 to 25~nm. This behavior is consistent with Eq.~(\ref{Eq:chi_DeltaE}), as the critical binding energy \(\Delta E_{\rm c}\) is insensitive to \(\ell_{\rm p}\) in the regime \(L \ll \ell_{\rm p}\). However, when \(\ell_{\rm p}\) becomes comparable to the arm length \(L\), \(\Delta E_{\rm c}\) decreases significantly. In this regime, the rigid-arm approximation breaks down, and the spatial distribution of sticky-end positions can no longer be assumed uniform within the thick shell, rendering our treatment invalid. Fortunately, for most DNA nanostars synthesized in experiments, the arm length (\(L \lesssim 10~\mathrm{nm}\)) is typically smaller than the persistence length (\(\ell_{\rm p} \approx 50~\mathrm{nm}\)) \cite{jeon2020sequence, sato2020sequence}, ensuring the validity of our theoretical framework. In the opposite limit, where \(L \gtrsim \ell_{\rm p}\), the conformational entropy loss upon binding becomes more pronounced because flexible arms favor unbound, fluctuating configurations. As a result, a stronger binding energy is required to drive condensation. Moreover, for sufficiently long arms, the probability of intramolecular (self-)binding increases, reducing the effective intermolecular interaction \(w_{11}\), since intermolecular binding then contributes less to lowering the free energy.

The above derivation considers only homogeneous DNA nanostars with identical sticky ends, but it can be readily extended to heterogeneous nanostars containing \(\alpha\) A-type sticky ends and \(\beta\) B-type sticky ends. Under the one-pair-binding assumption, the corresponding partition function becomes
\begin{align} \label{Eq:partition}
    \mathcal{Z}_{11} &\approx  \Omega_{\rm unbind} 
    + \alpha^2 \Omega_{\rm pair}
    e^{-\Delta E_{\rm AA} / (k_{\rm B}T)}
    + \beta^2 \Omega_{\rm pair}
    e^{-\Delta E_{\rm BB} / (k_{\rm B}T)}
    .
\end{align}
Here, \(\Delta E_{\rm AA}\) and \(\Delta E_{\rm BB}\) denote the binding energies of sticky ends~A and~B, respectively, and distinct types of sticky ends are assumed to interact purely repulsively. The configuration numbers \(\alpha^2\Omega_{\rm pair}\) and \(\beta^2\Omega_{\rm pair}\) represent, respectively, the number of combinatorial ways to bind two neighboring nanostars via A–A or B–B sticky-end pairs, together with the number of remaining configurations \(\Omega_{\rm pair}\) once a specific pair of DNA arms is selected [see Eq.~(\ref{Eq:Omega_pair})]. The Flory interaction parameter \(\chi\) is therefore given by
\begin{align} \label{Eq:chi_general}
     \chi &\approx
     \frac{z_{\rm eff}-2}{2} 
     \ln \frac{\mathcal{Z}_{11}}{\Omega_{\rm unbind}}
     \notag \\
     &= \frac{z_{\rm eff}-2}{2}
     \ln \left[
     1 + \alpha^2
     \exp\left(-\frac{\Delta E_{\rm AA}}{k_{\rm B}T} - 3\ln\overline L + C \right)
     + \beta^2
     \exp\left(-\frac{\Delta E_{\rm BB}}{k_{\rm B}T}  - 3\ln\overline L + C \right)
     \right]
     \notag \\
     &\approx \frac{z_{\rm eff}-2}{2}
     \left\{
     \ln \left[
     \alpha^2
     \exp\left(-\frac{\Delta E_{\rm AA}}{k_{\rm B}T} \right)
     + \beta^2
     \exp\left(-\frac{\Delta E_{\rm BB}}{k_{\rm B}T} \right)
     \right]
     - 3\ln\overline L + C
     \right\}.
\end{align}
The last line of Eq.~(\ref{Eq:chi_general}) assumes that the bound states dominate the partition function, which is valid once the system enters the phase-separated regime. In the special case where \(\Delta E_{\rm AA} = \Delta E_{\rm BB} = \Delta E\), Eq.~(\ref{Eq:chi_general}) reduces to
\begin{align} \label{Eq:chi_hetero}
     \chi \approx\frac{z_{\rm eff}-2}{2}
     \left[
     -\frac{\Delta E}{k_{\rm B}T}
     +\ln (\alpha^2+\beta^2)- 3 \ln\overline L + C (v_{\rm bind})
     \right].
\end{align}

We then consider systems containing \(n\) types of DNA nanostars, where each nanostar may be heterogeneous and carry an arbitrary combination of sticky ends of types A, B, C, and so on. For such \(n\)-component DNA-nanostar solutions, the Flory–Huggins free-energy density, \(f \equiv F/V\), is given by
\begin{align} \label{Eq:fx}
    \frac{f}{\rho_0 k_{\rm B}T} =
    \sum_{i=0}^n
    \frac{\phi_i}{M_{i,{\rm eff}}} \ln \phi_i
    +
    \frac{1}{2} 
    \sum_{i,j=0}^n
    \chi_{ij}
    \phi_i\phi_j
    ,
\end{align}
where subscripts \(i\) (\(i \neq 0\)) and “0” denote lattice sites occupied by nanostar type \(i\) and by pure solvent, respectively, and the volume fractions satisfy \(\sum_{i=0}^{n} \phi_i = 1\). In general, the polymerization factor satisfies \(M_{i,\rm eff} > 1\) for nanostar-occupied sites but \(M_{0,\rm eff} = 1\) for solvent-filled sites, as solvent molecules are isotropic and interact uniformly with their neighbors. For simplicity, we assume that the lattice-site number density \(\rho_0\) is identical for nanostars of different valence. This is a reasonable approximation because \(\rho_0\) varies only weakly with valence—for example, \(\rho_0 \approx 2.02\times10^{-3}\) and \(1.91\times10^{-3}~{\rm nm^{-3}}\) for \(Z=6\) and \(Z=4\), respectively, when \(L=5\) nm. This assumption effectively requires all nanostars to have comparable effective sizes; accordingly, we set \(L_i = L\) for all components below. Under these assumptions, the Flory interaction parameter between heterogeneous DNA nanostars and solvent lattice sites can be generalized as
\begin{align} \label{Eq:chi1chi2}
    \chi_i (i\neq 0) \equiv \chi_{i0}
    &\approx \frac{z_{i,\rm eff}-2}{2}
     \left\{
     \ln \left[
     \alpha_i^2
     \exp\left(-\frac{\Delta E_{\rm AA}}{k_{\rm B}T} \right)
     + \beta_i^2
     \exp\left(-\frac{\Delta E_{\rm BB}}{k_{\rm B}T} \right)
     + \dots
     \right]
     - 3\ln\overline L + C
     \right\}
     \notag \\
     &=\frac{z_{i,\rm eff}-2}{2}
     \left[
     -\frac{\Delta E}{k_{\rm B}T}
     +\ln (\alpha_i^2+\beta_i^2 + \dots)- 3 \ln\overline L + C (v_{\rm bind})
     \right],  {\quad \rm if \ } \Delta E_{\rm AA} = \Delta E_{\rm BB} = \dots,
\end{align}
where each component \(i\in\{1,2,\dots,n\}\) is specified by its sticky-end composition, denoted as “\(\alpha_i{\rm A}\,\beta_i{\rm B}\,\delta_i{\rm D}\dots\)” \((Z_i=\alpha_i+\beta_i+\delta_i+\dots)\), and characterized by an effective coordination number \(z_{i, \rm eff}\). The Flory interaction parameter between components \(i\) and \(j\) is defined by
\begin{align} \label{Eq:chi_ij}
    \chi_{ij} &= \frac{2(z_{ij,\rm eff}-2)w_{ij} - (z_{i,\rm eff}-2)w_{ii} - (z_{j, \rm eff}-2)w_{jj}}{2k_{\rm B}T}
    \notag\\
    & = \frac{(z_{ij, \rm eff}-2)w_{ij}- (z_{i,\rm eff}-2)w_{i0} - (z_{j, \rm eff}-2)w_{j0}}{k_{\rm B}T}
    + \chi_{i} + \chi_{j},
\end{align}
where \(z_{ij,\rm eff} \equiv (z_{i,\rm eff} + z_{j,\rm eff})/2\) denotes the effective coordination number between the two components. To evaluate the effective free energies \(w_{ij}\), \(w_{i0}\), and \(w_{j0}\), we consider the partition functions associated with two distinct neighboring DNA nanostars \(i\) and \(j\):
\begin{subequations}
\begin{align} 
    \mathcal{Z}_{ij} &\approx
    \Omega_{\rm unbind}^{(ij)} 
    + \Omega_{\rm bind,AA}^{(ij)} 
    e^{-\Delta E_{\rm AA} / (k_{\rm B}T)}
    + \Omega_{\rm bind,BB}^{(ij)} 
    e^{-\Delta E_{\rm BB} / (k_{\rm B}T)}
    + \dots, \label{Eq:Zij}
    \\
    \mathcal{Z}_{i0} &= \Omega_{\rm star}^{(i)},
    \\
    \mathcal{Z}_{j0} &= \Omega_{\rm star}^{(j)},
\end{align}    
\end{subequations}
where \(\Omega_{\rm unbind}^{(ij)}\) is the number of unbound configurations accessible to two nanostars of types \(i\) and \(j\), \(\Omega_{\rm bind,AA}^{(ij)}\) denotes the number of bound configurations formed through an A–A sticky-end pair (with analogous definitions for B–B, C–C, etc.), and \(\Omega_{\rm star}^{(i)}\) denotes the total number of configurations of nanostar \(i\). The corresponding state counts are
\begin{subequations}
\begin{align}
    \Omega_{\rm unbind}^{(ij)} &\approx \Omega_{\rm star}^{(i)} \Omega_{\rm star}^{(j)} = \Omega_{\rm arm}^2 \Omega_{\rm rest}^{(i)} \Omega_{\rm rest}^{(j)},
    \\
    \Omega_{\rm bind, AA}^{(ij)} &\approx
    \alpha_i \alpha_j \cdot 
    v_{\rm overlap}v_{\rm bind}/v_{\rm shell}^2
    \cdot \Omega_{\rm arm}^2\Omega_{\rm rest}^{(i)} \Omega_{\rm rest}^{(j)},
    \\
    \Omega_{\rm bind, BB}^{(ij)} &\approx
    \beta_i \beta_j \cdot 
    v_{\rm overlap}v_{\rm bind}/v_{\rm shell}^2
    \cdot \Omega_{\rm arm}^2\Omega_{\rm rest}^{(i)} \Omega_{\rm rest}^{(j)},
    \\
    \Omega_{\rm bind, CC}^{(ij)} &\approx \dots,
\end{align}    
\end{subequations}
where \(\Omega_{\rm rest}^{(i)}\) denotes the number of configurations of the remaining arms of nanostar \(i\) upon fixing one arm. Therefore, \(\chi_{ij}\) can be written as
\begin{align} \label{Eq:chi_ij_general}
    \chi_{ij} =\ & 
    -(z_{ij, \rm eff}-2)
     \ln \mathcal{Z}_{ij}
    + (z_{i, \rm eff}-2) \ln \mathcal{Z}_{i0}
    + (z_{j, \rm eff}-2) \ln \mathcal{Z}_{j0}
    + \chi_{i} + \chi_{j}
    \notag\\
    =\ & 
    -(z_{ij, \rm eff}-2)
     \ln \frac{\mathcal{Z}_{ij}}{\Omega_{\rm unbind}^{(ij)}}
    + \frac{z_{i, \rm eff} - z_{j, \rm eff}}{2}\ln\frac{\Omega_{\rm star}^{(i)}}{\Omega_{\rm star}^{(j)}}
    + \chi_{i} + \chi_{j}
    \notag\\
    \approx\ &
    -\frac{z_{i, \rm eff} + z_{j, \rm eff} -4}{2}
    \ln \left[
    1 +
    \alpha_i\alpha_j
    \exp\left(-\frac{\Delta E_{\rm AA}}{k_{\rm B}T} - 3\ln\overline L + C \right)
    + \beta_i\beta_j
    \exp\left(-\frac{\Delta E_{\rm AA}}{k_{\rm B}T} - 3\ln\overline L + C \right)
    + \dots
    \right]
    \notag\\
    \ & +
    \frac{z_{i,\rm eff}-2}{2}
    \left\{
    \ln \left[
    \alpha_i^2
    \exp\left(-\frac{\Delta E_{\rm AA}}{k_{\rm B}T} \right)
    + \beta_i^2
    \exp\left(-\frac{\Delta E_{\rm BB}}{k_{\rm B}T} \right)
    + \dots
    \right]
    - 3\ln\overline L + C
    \right\}
    \notag\\
    \ & +
    \frac{z_{j,\rm eff}-2}{2}
    \left\{
    \ln \left[
    \alpha_j^2
    \exp\left(-\frac{\Delta E_{\rm AA}}{k_{\rm B}T} \right)
    + \beta_j^2
    \exp\left(-\frac{\Delta E_{\rm BB}}{k_{\rm B}T} \right)
    + \dots
    \right]
    - 3\ln\overline L + C
    \right\}
    \notag\\
    \approx \ &
    \frac{z_{i,\rm eff}-2}{2}
    \ln \frac{
    \alpha_i^2
    \exp\left(-\frac{\Delta E_{\rm AA}}{k_{\rm B}T} \right)
    + \beta_i^2
    \exp\left(-\frac{\Delta E_{\rm BB}}{k_{\rm B}T} \right)
    + \dots}
    {\alpha_i\alpha_j
    \exp\left(-\frac{\Delta E_{\rm AA}}{k_{\rm B}T} \right)
    + \beta_i\beta_j
    \exp\left(-\frac{\Delta E_{\rm BB}}{k_{\rm B}T} \right)
    + \dots}
    \notag \\
    \ &+
    \frac{z_{j,\rm eff}-2}{2}
    \ln \frac{
    \alpha_j^2
    \exp\left(-\frac{\Delta E_{\rm AA}}{k_{\rm B}T} \right)
    + \beta_j^2
    \exp\left(-\frac{\Delta E_{\rm BB}}{k_{\rm B}T} \right)
    + \dots}
    {\alpha_i\alpha_j
    \exp\left(-\frac{\Delta E_{\rm AA}}{k_{\rm B}T} \right)
    + \beta_i\beta_j
    \exp\left(-\frac{\Delta E_{\rm BB}}{k_{\rm B}T} \right)
    + \dots}
    \notag\\
    =\ & \frac{z_{i,\rm eff}-2}{2}
    \ln \frac{\alpha_i^2 + \beta_i^2 + \dots}
    {\alpha_i\alpha_j + \beta_i\beta_j + \dots}
    +
    \frac{z_{j,\rm eff}-2}{2}
    \ln \frac{\alpha_j^2 + \beta_j^2 + \dots}
    {\alpha_i\alpha_j + \beta_i\beta_j + \dots}
    ,  {\quad \rm if \ } \Delta E_{\rm AA} = \Delta E_{\rm BB} = \dots \ ,
\end{align}
where we have neglected the constant term \((z_{i, \rm eff} - z_{j, \rm eff})/2 \cdot \ln(\Omega_{\rm star}^{(i)}/\Omega_{\rm star}^{(j)})\), which is independent of the sticky-end distributions \(\{\alpha_i, \beta_i, \dots\}\). This term is expected to be small because the difference in \(z_{i, \rm eff}\) is minor (see Table~\ref{table:FH}), while \(\Omega_{\rm star}^{(i)}\) and \(\Omega_{\rm star}^{(j)}\) should be comparable in magnitude under strong steric repulsion between DNA arms. From the third-to-last equality in Eq.~(\ref{Eq:chi_ij_general}), we assume that the total binding configuration number, \(\Omega_{\rm bind}^{(ij)} = \Omega_{\rm bind,AA}^{(ij)} + \Omega_{\rm bind,BB}^{(ij)} + \dots\), is nonzero and dominates the partition function in the phase-separated regime, which requires \(\alpha_i\alpha_j + \beta_i\beta_j + \dots \neq 0\). If \(\Omega_{\rm bind}^{(ij)} = 0\), then only the unbound term \(\Omega_{\rm unbind}^{(ij)}\) contributes to the partition function [Eq.~(\ref{Eq:Zij})], giving \(\chi_{ij} \approx \chi_i + \chi_j\). Equation~(\ref{Eq:chi_ij_general}) also inherently ensures that \(\chi_{ii} = 0\), as required in the Flory–Huggins theory.

It should be noted that our generalized Flory–Huggins theory accounts only for pairwise interactions between DNA nanostars. Consequently, for a heterogeneous nanostar with sticky-end composition “\(\alpha_i{\rm A}\beta_i{\rm B}c_i{\rm C}\dots\)” (\(Z_i=\alpha_i+\beta_i+c_i+\dots\)), the predicted phase behavior depends only on the numbers of each sticky-end type, not on their specific arrangement along the molecule. For example, the variants “AAAABB,” “AAABAB,” and “AABAAB” all share the same composition and therefore possess identical theoretical phase behavior. In practice, these variants also behave similarly, as DNA nanostars typically have semiflexible junctions that allow different arm orderings to produce comparable conformational statistics. As shown in Fig.~\ref{sequence}, the interfacial energies of condensates composed of the heterogeneous nanostars “AAAABB,” “AAABAB,” and “AABAAB” are nearly identical.

\newpage

\renewcommand{\thetable}{S\arabic{table}}
\begin{table}[htbp]
    \centering
    \includegraphics[width=0.95\linewidth]{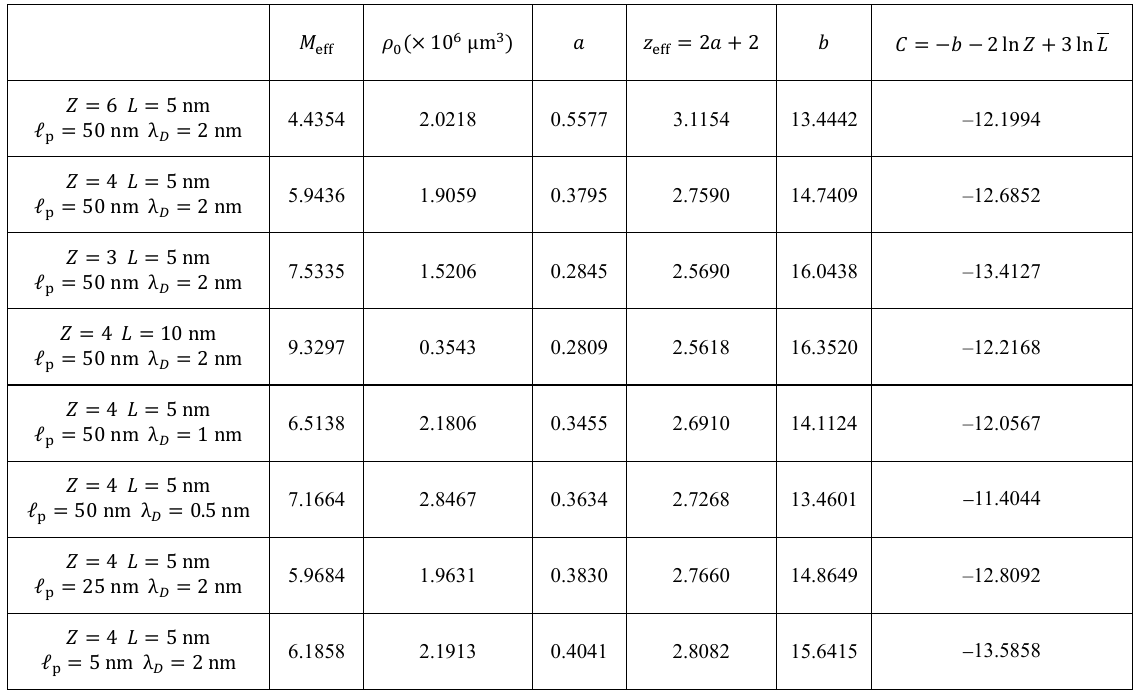}
    \caption{Summary of the fitting parameters in the Flory–Huggins theory, where \(\chi \equiv -a \big[\Delta E/(k_{\rm B}T) + b\big]\). The fitted parameters \(a\) and \(b\) correspond to \(z_{\rm eff}\) and \(C\) in Eq.~(\ref{Eq:chi_DeltaE}).}
    \label{table:FH}
\end{table}

\renewcommand{\thefigure}{S\arabic{figure}}
\begin{figure*}[!hbt]
	\centering
	\includegraphics[width=0.5\linewidth]{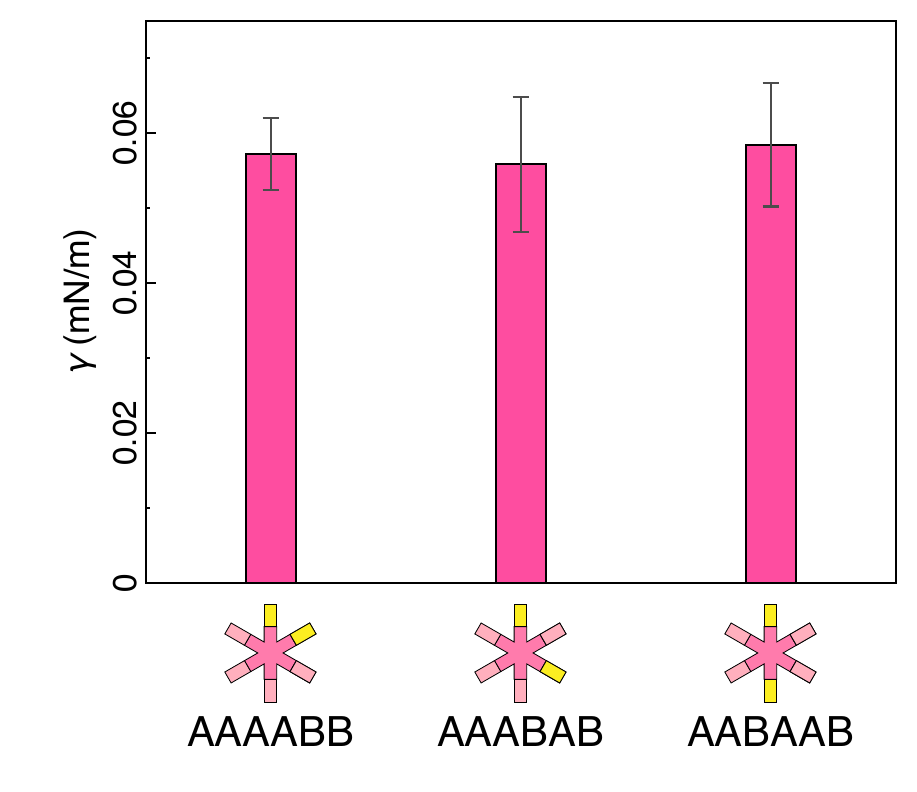}
	\caption{Interfacial energy of heterogeneous nanostars with different sticky-end arrangements, where \(\Delta E_{\rm AA} = \Delta E_{\rm BB} = -19 k_{\rm B}T\). The error bars for each result represent the SEM calculated from five independent simulations.}
	\label{sequence}
\end{figure*}

\newpage
\section{Derivation of the interfacial energy in a multi-component system} \label{sec3}

Based on the generalized Flory–Huggins theory for DNA nanostars, the interfacial energy between two phases \(I\) and \(J\) can be evaluated by considering a one-dimensional system composed of \(n\) types of DNA nanostars. The volume fraction \(\phi_i(x)\) is assumed to be uniform in the other two spatial dimensions, and the boundary conditions are specified as \(\lim_{x \to -\infty}\phi_i(x) = \phi_i^I\) and \(\lim_{x \to +\infty}\phi_i(x) = \phi_i^J\), where \(\phi_i^I\) is the equilibrium volume fraction of species \(i\) in phase \(I\). Here \(i \in \{0,1,2,\dots,n\}\), where \(i=0\) denotes lattice sites occupied by solvent and \(i>0\) corresponds to DNA nanostars. At equilibrium, a unique interface with a well-defined interfacial energy forms at finite \(x\), corresponding to the minimum of the free-energy functional. Following the classic approach of Cahn and Hilliard \cite{cahn1958free}, the total free-energy functional for this one-dimensional system is given by
\begin{equation}
    \frac{F[\{\phi_i(x)\}, \mu(x)]}{\rho_0 A k_{\rm B}T} = \int {\rm d}x \left[
    \sum_{i=0}^n \frac{\phi_i}{M_{{\rm eff},i}} \ln \phi_i
    +
    \frac{1}{2} \sum_{i,j=0}^n \phi_i\chi_{ij}\phi_j
    -
    \frac{\lambda^2}{2} \sum_{i,j=0}^n \phi'_i \chi_{ij} \phi'_j
    - \mu \left(
    \sum_{i=0}^n \phi_i - 1
    \right)
    \right].
\end{equation}
Here, \(\rho_0\) is the lattice-site number density, which is assumed to be identical for all species; \(A\) is the cross-sectional area; \(\phi_i' \equiv {\rm d}\phi_i/{\rm d}x\) denotes the spatial derivative; and \(\mu(x)\) is the Lagrange multiplier that enforces the local incompressibility condition \(\sum_{i=0}^n \phi_i = 1\), serving as a generalized local pressure field. The factor \(\lambda\) represents a characteristic length scale that sets the interfacial width, while the gradient term \(\lambda^2 \phi_i' \chi_{ij} \phi_j'\) accounts for the interfacial energy arising from the penalty on spatial inhomogeneity. In contrast to this gradient contribution, the homogeneous (dimensionless) part of the free-energy density is given by
\begin{equation} \label{Eq:f_homo}
    \overline f(\phi_i) = 
    \sum_{i=0}^n \frac{\phi_i}{M_{{\rm eff},i}} \ln \phi_i
    +
    \frac{1}{2} \sum_{i,j=0}^n \phi_i\chi_{ij}\phi_j
    .
\end{equation}
The interfacial energy arises from two sources: the local composition deviates from its equilibrium values in the coexisting bulk phases, and the concentration varies spatially across the interface. Assuming that the volume fraction \(\phi_i(x)\) at the interface lies between the equilibrium values \(\phi_i^I\) and \(\phi_i^J\), the corresponding excess free-energy density associated with these non-equilibrium compositions is given by
\begin{align} \label{Eq:Delta_f}
    \Delta \overline f(\phi_i) &=
    \overline f(\phi_i)
    - \eta \overline f(\phi_i^I)
    - (1-\eta) \overline f(\phi_i^J),
    \\
    \phi_i &= \eta \phi_i^I + (1-\eta) \phi_i^J, \label{Eq:eta}
\end{align}
where \(\eta \in [0,1]\) is an interpolation factor that characterizes the local composition across the interface. The interfacial energy between Phases \(I\) and \(J\) is therefore defined as the sum of this excess free-energy density and the gradient contribution, namely
\begin{equation}
    \frac{\gamma_{IJ}[\{\phi_i(x)\}, \mu(x)]}{\rho_0 k_{\rm B}T} = \int {\rm d}x \left[
    \Delta \overline f
    -
    \frac{\lambda^2}{2} \sum_{i,j=0}^n \phi'_i \chi_{ij} \phi'_j
    - \mu \left(
    \sum_{i=0}^n \phi_i - 1
    \right)
    \right].
\end{equation}
At equilibrium, the interfacial profile minimizes this functional with respect to all admissible fields, so that \(\gamma_{IJ}\) attains its minimum value for the equilibrium interface. This requires that the variations of the functional with respect to all relevant fields vanish, leading to
\begin{subequations}
\begin{align}
    \frac{\delta\gamma_{IJ}}{\delta \phi_i} &\propto
    \frac{\partial \Delta \overline f}{\partial \phi_i}
    + \lambda^2 \sum_{\substack{j=0\\j\neq i}}^n \chi_{ij} \phi''_j
    - \mu = 0, \quad i \in \{0, 1, \dots, n \}, \label{Eq:delta gamma}\\
    \frac{\delta\gamma_{IJ}}{\delta \mu} &\propto
    1-\sum_{i=0}^n \phi_i = 0.
\end{align}    
\end{subequations}
In Eq.~(\ref{Eq:delta gamma}), we can subtract the equation for \(i=0\) from that for each \(i \neq 0\) to eliminate the Lagrange multiplier \(\mu\), yielding
\begin{align}
    0=&\frac{\partial \Delta \overline f}{\partial \phi_i}
    - \frac{\partial \Delta \overline f}{\partial \phi_0}
    + \lambda^2 \sum_{\substack{j=0\\j\neq i}}^n \chi_{ij} \phi''_j
    - \lambda^2 \sum_{j=1}^n \chi_{0j} \phi''_j \notag\\
    =
    &\frac{\partial \Delta \overline f}{\partial \phi_i}
    - \frac{\partial \Delta \overline f}{\partial \phi_0}
    + \lambda^2 \sum_{j=0}^n (\chi_{ij} - \chi_{0j}) \phi''_j \notag\\
    =
    &\frac{\partial \Delta \overline f}{\partial \phi_i}
    - \frac{\partial \Delta \overline f}{\partial \phi_0}
    + \lambda^2 \sum_{j=1}^n (\chi_{ij} - \chi_{0j}) \phi''_j
    + \lambda^2 (\chi_{i0} - \chi_{00}) \phi''_0 \notag\\
    =
    &\frac{\partial \Delta \overline f}{\partial \phi_i}
    - \frac{\partial \Delta \overline f}{\partial \phi_0}
    + \lambda^2 \sum_{j=1}^n (\chi_{ij} - \chi_{0j}) \phi''_j
    - \lambda^2 \chi_{0i} \sum_{j=1}^n \phi''_j \notag\\
    =
    &\frac{\partial \Delta \overline f}{\partial \phi_i}
    - \frac{\partial \Delta \overline f}{\partial \phi_0}
    + \lambda^2 \sum_{j=1}^n (\chi_{ij} - \chi_{0i} - \chi_{0j}) \phi''_j
    , \quad i \in \{1, \dots, n \}, \label{Eq:i1N}
\end{align}
where the summation conditions \(j \neq i\) and \(j \neq 0\) in the first line can be omitted because \(\chi_{ii} = 0\), and the identity \(\sum_{i=0}^n \phi_i = 1\) is used in the fourth line. Multiplying Eq.~(\ref{Eq:i1N}) by \(\phi_i' \equiv {\rm d}\phi_i/{\rm d}x\) and summing over \(i\) from 1 to \(n\) yields
\begin{align} \label{Eq:1}
    0=&\sum_{i=1}^n
    \frac{\partial \Delta \overline f}{\partial \phi_i}
    \frac{{\rm d} \phi_i}{{\rm d}x}
    - \sum_{i=1}^n
    \frac{\partial \Delta \overline f}{\partial \phi_0}
    \frac{{\rm d} \phi_i}{{\rm d}x}
    + \lambda^2 \sum_{i,j=1}^n (\chi_{ij} - \chi_{0i} - \chi_{0j}) \phi_i'\phi''_j \notag \\
    =
    &\sum_{i=1}^n
    \frac{\partial \Delta \overline f}{\partial \phi_i}
    \frac{{\rm d} \phi_i}{{\rm d}x}
    +
    \frac{\partial \Delta \overline f}{\partial \phi_0}
    \frac{{\rm d} \phi_0}{{\rm d}x}
    + \frac{{\rm d}}{{\rm d}x}
    \left[
    \frac{\lambda^2}{2} \sum_{i,j=0}^n (\chi_{ij} - \chi_{0i} - \chi_{0j}) \phi_i'\phi'_j
    \right]
    \notag \\
    =
    &\frac{{\rm d}\Delta \overline f}{{\rm d}x}
    +\frac{{\rm d}}{{\rm d}x}
    \left[
    \frac{\lambda^2}{2} \sum_{i,j=0}^n
    \phi_i' \chi_{ij} \phi'_j
    - \lambda^2 \sum_{i,j=0}^n
    \phi_i' \chi_{0i} \phi'_j
    \right] \notag \\
    =
    &\frac{{\rm d}}{{\rm d}x}
    \left[
    \Delta \overline f
    +
    \frac{\lambda^2}{2} \sum_{i,j=0}^n \phi_i' \chi_{ij} \phi'_j
    - \lambda^2 \sum_{i=0}^n
    \phi_i' \chi_{0i} \sum_{j=0}^n \phi'_j
    \right] \notag \\
    =
    &\frac{{\rm d}}{{\rm d}x}
    \left[
    \Delta \overline f
    +
    \frac{\lambda^2}{2} \sum_{i,j=0}^n \phi_i' \chi_{ij} \phi'_j
    \right]
    ,
\end{align}
where we have used \(\sum_{j=0}^n \phi_j' = 0\). Equation~(\ref{Eq:1}) indicates that
\begin{equation}
    \Delta \overline f
    +
    \frac{\lambda^2}{2} \sum_{i,j=0}^n \phi_i' \chi_{ij} \phi'_j
    = \rm constant = 0,
    \label{Eq:lagrangian}
\end{equation}
since the constant term vanishes at the homogeneous boundaries. Therefore, the interfacial energy is given by
\begin{equation}
    \frac{\gamma_{IJ}[\{\phi_i(x)\}]}{\rho_0 k_{\rm B}T} = 2 \int {\rm d}x
    \Delta \overline f, \label{Eq:gamma_ab}
\end{equation}
which depends only on the excess free-energy density \(\Delta \overline f(x)\). From Eq.~(\ref{Eq:lagrangian}), \(\Delta \overline f(x)\) can be written as
\begin{equation}
    \Delta \overline f
    =
    - \frac{\lambda^2}{2}  \phi_i' \chi_{ij} \phi'_j
    =
    - \frac{\lambda^2}{2} |\vec\phi'|^2
    \hat t_i \chi_{ij} \hat t_j.
\end{equation}
Here, \(\vec{\phi} \equiv [\phi_i]\) denotes the vector of volume fractions in the \((n+1)\)-dimensional composition space; \(\hat t \equiv \vec{\phi}'/|\vec{\phi}'|\) is the unit tangent to the trajectory \(\vec{\phi}(x)\) parameterized by \(x\); and \(|\vec{\phi}'| \equiv \sqrt{\phi'_i \phi'_i}\) is the Euclidean norm of \(\vec{\phi}'\). The Einstein summation convention is used henceforth. Because \(\Delta \overline{f} \ge 0\), we have \(\hat{t}_i \chi_{ij} \hat{t}_j \le 0\). Consequently, \(|\vec{\phi}'|\) is given by
\begin{equation}
    |\vec\phi'| =
    \frac{1}{\lambda}
    \sqrt{-\frac{2\Delta \overline f}{\hat t_i \chi_{ij} \hat t_j}}.
\end{equation}
We now change variables from \(x\) to the arc length in \(\vec{\phi}\)-space,
\begin{equation}
    |{\rm d}\vec\phi| = |\vec\phi'| {\rm d}x \Rightarrow {\rm d}x = \frac{|{\rm d}\vec\phi|}{|\vec\phi'|}.
\end{equation}
Thus, Eq.~(\ref{Eq:gamma_ab}) can be rewritten as
\begin{equation}
    \frac{\gamma_{IJ}[\phi_i(x)]}{\rho_0 k_{\rm B}T}
    = 2 \int_{\vec\phi^I}^{\vec\phi^J}
    |{\rm d}\vec\phi|
    \frac{\Delta \overline f}{|\vec\phi'|}
    = 2 \lambda \int_{\vec\phi^I}^{\vec\phi^J}
    |{\rm d}\vec \phi|
    \sqrt{-\frac{1}{2}\hat t_i \chi_{ij} \hat t_j\Delta\overline f}.
\end{equation}
Across the interface between phases \(I\) and \(J\), we assume that the trajectory of \(\vec{\phi}(x)\) is a straight line connecting \(\vec{\phi}^I\) and \(\vec{\phi}^J\), implying that
\begin{equation}
    \hat t = \frac{\vec\phi^J - \vec\phi^I}{|\vec\phi^J - \vec\phi^I|}.
\end{equation}
Therefore, the interfacial energy is given by
\begin{equation} \label{Eq:gamma_IJ}
    \frac{\gamma_{IJ}[\phi_i(x)]}{\rho_0 k_{\rm B}T}
    = 2 \lambda \int_{\vec\phi^I}^{\vec\phi^J}
    \frac{|{\rm d}\vec \phi|}{|\vec\phi^J - \vec\phi^I|}
    \sqrt{-\frac{1}{2} (\phi_i^I - \phi_i^J) \chi_{ij} (\phi_j^I - \phi_j^J) \Delta\overline f}
    \equiv 2 \lambda \int_{0}^{1}
    {\rm d}\eta
    \sqrt{K_{IJ} \Delta\overline f},
\end{equation}
with an effective coupling between Phases \(I\) and \(J\) defined by
\begin{equation}
    K_{IJ} \equiv
    -\frac{1}{2} \sum_{i, j=0}^n (\phi_i^I - \phi_i^J) \chi_{ij} (\phi_j^I - \phi_j^J)
    .
\end{equation}
In a binary system with \(n = 1\), Eq.~(\ref{Eq:gamma_IJ}) reduces to
\begin{align} \label{Eq:gamma}
    \gamma = 2\lambda \rho_0 k_{\rm B}T\int_{\phi_l}^{\phi_h} {\rm d}\phi \sqrt{\chi \Delta \overline f(\phi)},
\end{align}
where \(\phi_l\) and \(\phi_h\) denote the equilibrium volume fractions of the dilute and dense phases, respectively.

In the limit \(\chi_{ij} \gg 1\) for all \(i\) and \(j\), we may label Phases \(I\) and \(J\) such that they are predominantly composed of species \(I\) and \(J\), respectively; that is, \(\phi_i^I \approx \delta_{iI}\) and \(\phi_i^J \approx \delta_{iJ}\), where \(\delta_{ij}\) is the Kronecker delta. In this regime, the homogeneous free energy in Eq.~(\ref{Eq:f_homo}) is dominated by the interaction term \(\phi_i \chi_{ij} \phi_j\), and the excess free-energy density in Eq.~(\ref{Eq:Delta_f}) can be approximated as \(\Delta \overline f \approx \eta(1-\eta)\,\chi_{I\!J}\). The effective coupling term also satisfies \(K_{I\!J} \approx \chi_{I\!J}\). Consequently, the interfacial energy becomes
\begin{align} \label{Eq:gamma_approx}
    \gamma_{I\!J}
    \approx \frac{\pi}{4}\lambda \rho_0 k_{\rm B}T \chi_{I\!J}
    ,
\end{align}
which is proportional to \(\chi_{I\!J}\) and independent of the other components of the Flory interaction matrix \([\chi_{ij}]\). In this work, the interfacial length scale \(\lambda\) is assumed to be on the order of the lattice spacing and is written as \(\lambda = \mathcal{C}\,\rho_0^{-1/3}\), where \(\mathcal{C}\) is a fitting constant with a value of approximately 0.64. The interfacial energies predicted by Eq.~(\ref{Eq:gamma_IJ}) are shown as solid lines in Fig.~2b,~d for various conditions, demonstrating good agreement with the simulation results. Equations~(\ref{Eq:gamma_approx}) and~(\ref{Eq:chi_DeltaE}) together show that the prefactor \(\lambda \rho_0 (z_{\rm eff}-2)\) controls the overall slope of the interfacial energy as a function of the binding energy \(\Delta E\) in Fig.~2b,~d at large \(\Delta E\), where the lattice-site density \(\rho_0\) and the effective lattice coordination number \(z_{\rm eff}\) are determined by fitting the theoretical binodal to the simulation data. The factor \(\mathcal{C}\) may vary across systems because the coordination number of the generalized lattice can change; however, a single value \(\mathcal{C} = 0.64\) is used for all cases in Fig.~2b,~d, accounting for the small discrepancies between theory and simulation.

\newpage

\section{Details of diffusive phase-field simulations} \label{sec2}

Although molecular dynamics simulations are used for most of this work, certain arguments become significantly simpler—or require verification—when examined through continuum phase-field simulations. Following Cahn and Hilliard \cite{cahn1958free}, the total free energy \(F\) of a one-dimensional system composed of \(n\) types of DNA nanostars can be written as
\begin{align}
    \frac{F}{\rho_0 A k_{\rm B}T} = \int {\rm d}x \left[
    \sum_{i=0}^n\frac{\phi_i}{M_{i, \rm eff}}\ln \phi_i
    + \frac{1}{2}\sum_{i,j=0}^n
    \chi_{ij}\left(\phi_i\phi_j
    - \lambda^2\nabla\phi_i\cdot \nabla\phi_j\right)
    \right],
\end{align}
where \(\rho_0\) is the lattice-site number density (assumed uniform across all species), \(A\) is the cross-sectional area, and \(M_{i,{\rm eff}}\) denotes the effective polymerization factor for each species (with \(M_{0,{\rm eff}} = 1\) for the solvent). The field \(\phi_i(x)\) represents the volume fraction of species \(i\), where \(i=0\) corresponds to solvent and \(i>0\) corresponds to DNA nanostars, and \(\lambda\) is a characteristic length scale that determines the interfacial width. Because the volume fractions satisfy the local incompressibility condition \(\sum_{i=0}^n \phi_i = 1\), only \(n\) independent degrees of freedom exist in composition space. We therefore omit \(\phi_0\) in the subsequent analysis and focus on the concentration fields of the DNA nanostar species. The evolution of each component follows Model~B dynamics \cite{hohenberg1977theory}, i.e., conserved concentration dynamics, given by
\begin{align} \label{Eq:evo}
    \frac{\partial \phi_i}{\partial t} &= \nabla \cdot \left[
    \sum_{j=1}^n M_{ij} \nabla \mu_j
    \right], \quad i \in \{1, \dots, n\},
    \\
    \mu_i &= \frac{\delta F}{\delta \phi_i},
    \quad i \in \{1, \dots, n\},
    \notag\\&= \rho_0 A k_{\rm B}T \left[
    \frac{1}{M_{i, \rm eff}} -1 + \chi_{i0} + \frac{1}{M_{i,\rm eff}}\ln \phi_i -\ln\left(1 - \sum_{j=1}^n\phi_j\right)+ \sum_{j=1}^n (\chi_{ij} - \chi_{i0} - \chi_{j0}) (1 + \lambda^2\nabla^2) \phi_j
    \right],
\end{align}
where \(\mu_i\) is the one-dimensional chemical potential density of component \(i\) (\(i>0\)), and \(M_{ij}\) is the Onsager mobility matrix. For simplicity, we take \(M_{ij}\) to be diagonal and spatially uniform, i.e., \(M_{ij} = \delta_{ij}\lambda^2/(\rho_0 A k_{\rm B}T t_0)\) , where \(\delta_{ij}\) is the Kronecker delta and \(t_0\) denotes the unit of time. Since our interest is solely in the equilibrium state, the specific choice of \(M_{ij}\) does not affect the final equilibrium solution and is made primarily for computational efficiency. We solve the evolution equation (\ref{Eq:evo}) using an implicit–explicit scheme \cite{ascher1997implicit}, with additional details provided in Ref.~\cite{mao2019phase}.

As shown in Fig.~\ref{figs phase diagram}, we verify the theoretical predictions and molecular dynamics results for the interfacial energies in Fig.~4c using phase-field simulations. The continuum phase-field simulations are in good agreement with theory for both \(\gamma_2\) and \(\gamma_{12}\) [crosses in Fig.~\ref{figs phase diagram}a versus solid lines from Eq.~(\ref{Eq:gamma})], but yield a smaller value for \(\gamma_1\). This discrepancy arises because the theoretical calculation assumes that the composition change across an interface follows a straight-line path in composition space [Eq.~(\ref{Eq:eta})], whereas the true equilibrium interfacial paths are generally more complex. Figure~\ref{figs phase diagram}b–i shows the spatial volume-fraction profiles together with their trajectories in the ternary phase diagram, demonstrating that all interfacial paths deviate from the straight-line approximation. The deviations of the interfacial composition paths for the 1–2 and 2–0 interfaces are relatively small, allowing the theoretical predictions for \(\gamma_{12}\) and \(\gamma_2\) to remain accurate.

For the interface between phases~1 and~0, component~2 accumulates strongly at the interface (Fig.~\ref{figs phase diagram}b–d,~g), acting as a surfactant or forming an extended layer of nearly homogeneous composition. In the phase-field simulations, diffusion is sufficiently fast that the interfacial profile rapidly relaxes to its equilibrium configuration. Consequently, when the bare interfacial energy satisfies \(\gamma_1 > \gamma_{12} + \gamma_2\), the direct 1–0 interface is replaced by a wetted composite 1–2–0 interface, and the measured interfacial energy saturates at \(\gamma_1 \approx \gamma_{12} + \gamma_2\). This behavior is observed for the “4A2B+4B” system (Fig.~\ref{figs phase diagram}a), where the interface between phases~1 and~0 is fully wetted by phase~2 (Fig.~\ref{figs phase diagram}c,~g). Phase-field simulations accordingly cannot be used to extract the metastable (non-wetted) value of \(\gamma_1\) in this regime, whereas molecular dynamics simulations continue to probe the bare interfacial energy, consistent with the theoretical trend (Fig.~\ref{figs phase diagram}a). In contrast, for the “5A1B+4B’’ system, the 1–0 interface remains stable because \(\gamma_1 < \gamma_2 + \gamma_{12}\), and the phase-field simulations thus yield a value of \(\gamma_1\) comparable to that obtained from molecular dynamics simulations (pink bar in Fig.~\ref{figs phase diagram}a). Taken together, comparison across the three approaches indicates that the theoretical framework reliably predicts the qualitative morphological trends associated with varying sticky-end compositions, although the precise location of the critical condition \(\gamma_1 = \gamma_2 + \gamma_{12}\) shows modest deviations among the methods.

\renewcommand{\thefigure}{S\arabic{figure}}
\begin{figure*}[!htb]
	\centering
	\includegraphics[width=1\linewidth]{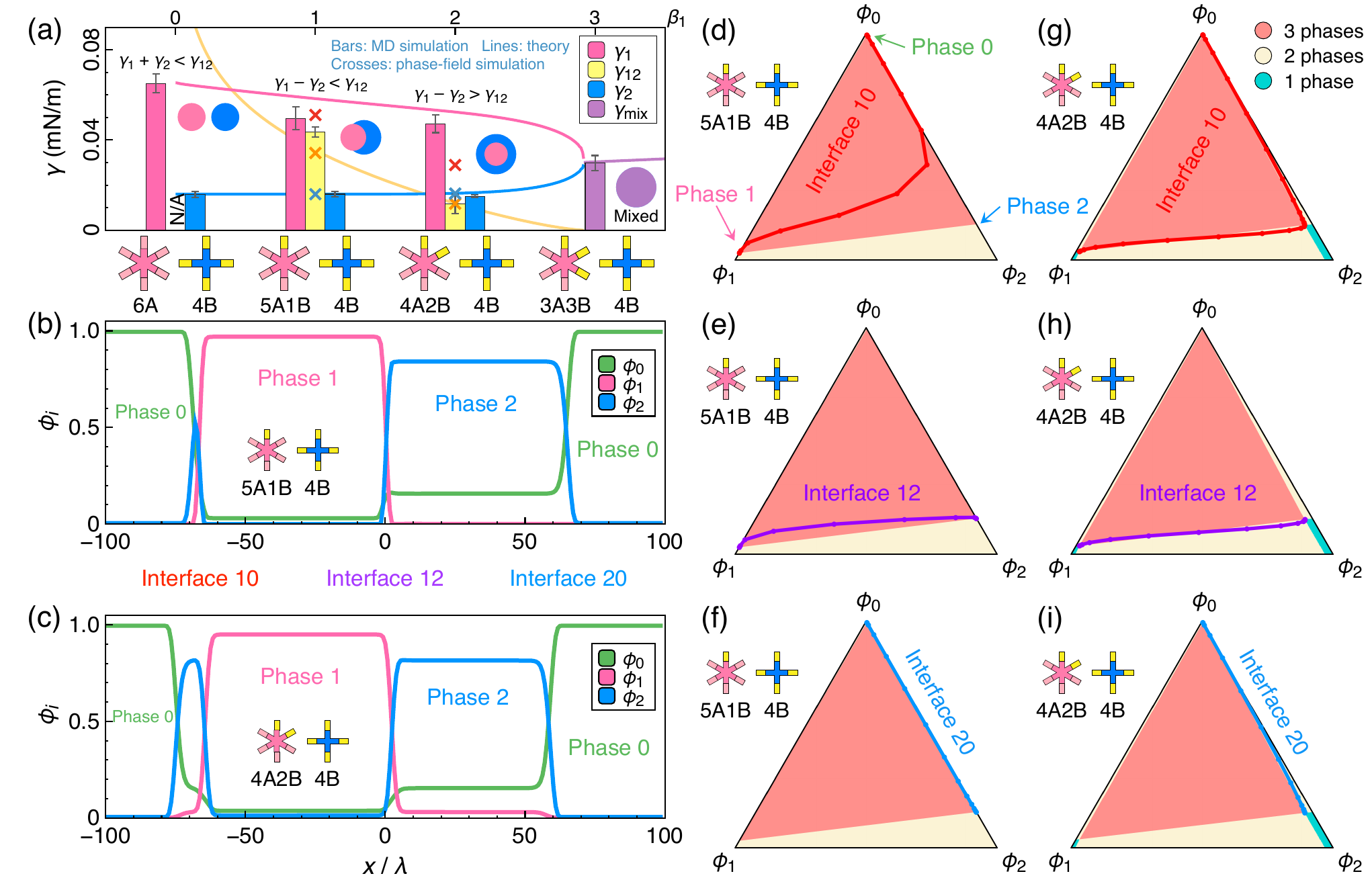}
	\caption{
    (a) Comparison of the interfacial energies shown in Fig.~4c with the phase-field simulation results (crosses).  
    (b,~c) Volume fraction profiles along the spatial coordinate for (b) “5A1B+4B” and (c) “4A2B+4B” systems.  
    (d–i) Volume fraction trajectories across each interface, plotted in the ternary phase diagram, for (d–f) “5A1B+4B” and (g–i) “4A2B+4B” systems.
    }
	\label{figs phase diagram}
\end{figure*}

For the simulations involving systems with crosslinkers (3+1 species; Fig.~6e–h), we set \(M_{{\rm eff},i} = 1\) for all DNA nanostar components. When the number of species, the values of \(M_{{\rm eff},i}\), and the interaction parameters \(\chi_{ij}\) are all large, each phase becomes dominated by a single component, while the concentrations of the remaining components drop to extremely low levels. Under such conditions, the timestep required for numerical stability becomes prohibitively small, causing the simulation to stall. As an alternative, we fix \(M_{{\rm eff},i} = 1\) and adjust \(\chi_{ij}\) to reproduce the same interfacial energies between each pair of phases. As shown in Fig.~\ref{Meff}, we compute the interfacial energies for binary systems with varying \(M_{\rm eff}\) and \(\chi\). The results indicate that changing \(M_{\rm eff}\) simply shifts the interfacial-energy curve. Thus, when we choose \(M_{\rm eff}=1\) in the simulations, we compensate by increasing \(\chi\) so as to maintain the desired interfacial energies. The interaction parameters used in the phase-field simulations in Fig.~6 are \(\chi_1 = 4\), \(\chi_2 = 2.6\), \(\chi_3 = 2.35\), \(\chi_{12} = 6.6\), \(\chi_{13} = 1.46\), and \(\chi_{23} = 1.26\), where \(\chi_{ij} = \chi_{ji}\) and \(\chi_{i} \equiv \chi_{i0}\).

\renewcommand{\thefigure}{S\arabic{figure}}
\begin{figure*}[!htb]
	\centering
	\includegraphics[width=0.5\linewidth]{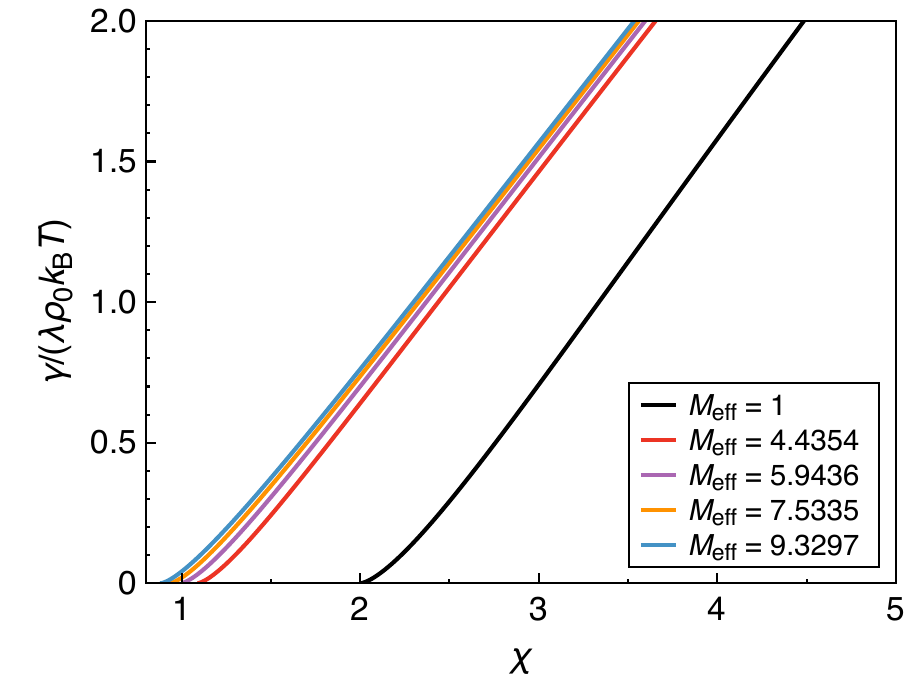}
	\caption{Interfacial energy as a function of the Flory interaction parameter \(\chi\) in a binary system, shown for different values of the effective polymerization factor \(M_{\rm eff}\).}
	\label{Meff}
\end{figure*}

\newpage
\bibliography{nanostar_ref}

\newpage
\renewcommand{\thefigure}{S\arabic{figure}}
\begin{figure*}[!htb]
	\centering
	\includegraphics[width=0.5\linewidth]{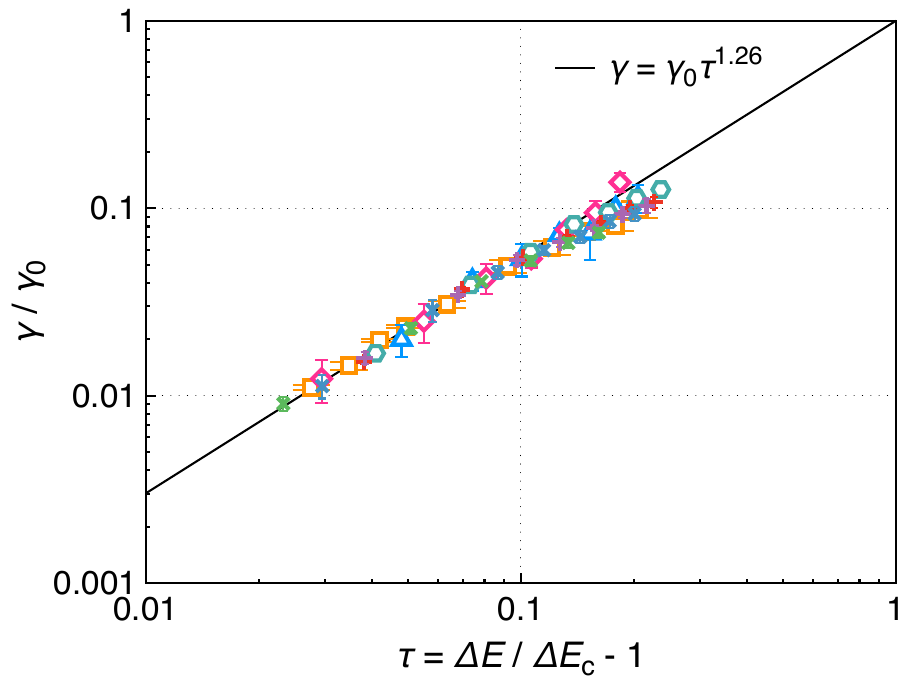}
	\caption{Power-law scaling of the interfacial energy near the critical point, where \(\tau = \Delta E / \Delta E_{\rm c} - 1\) and \(\gamma_0\) are fitting parameters. The fitting curve is accurate for \(\tau < 0.1\). The point labels match those in Fig.~2. Error bars represent the SEM computed from five independent simulations.}
	\label{figs1}
\end{figure*}

\renewcommand{\thefigure}{S\arabic{figure}}
\begin{figure*}[!htb]
	\centering
	\includegraphics[width=0.8\linewidth]{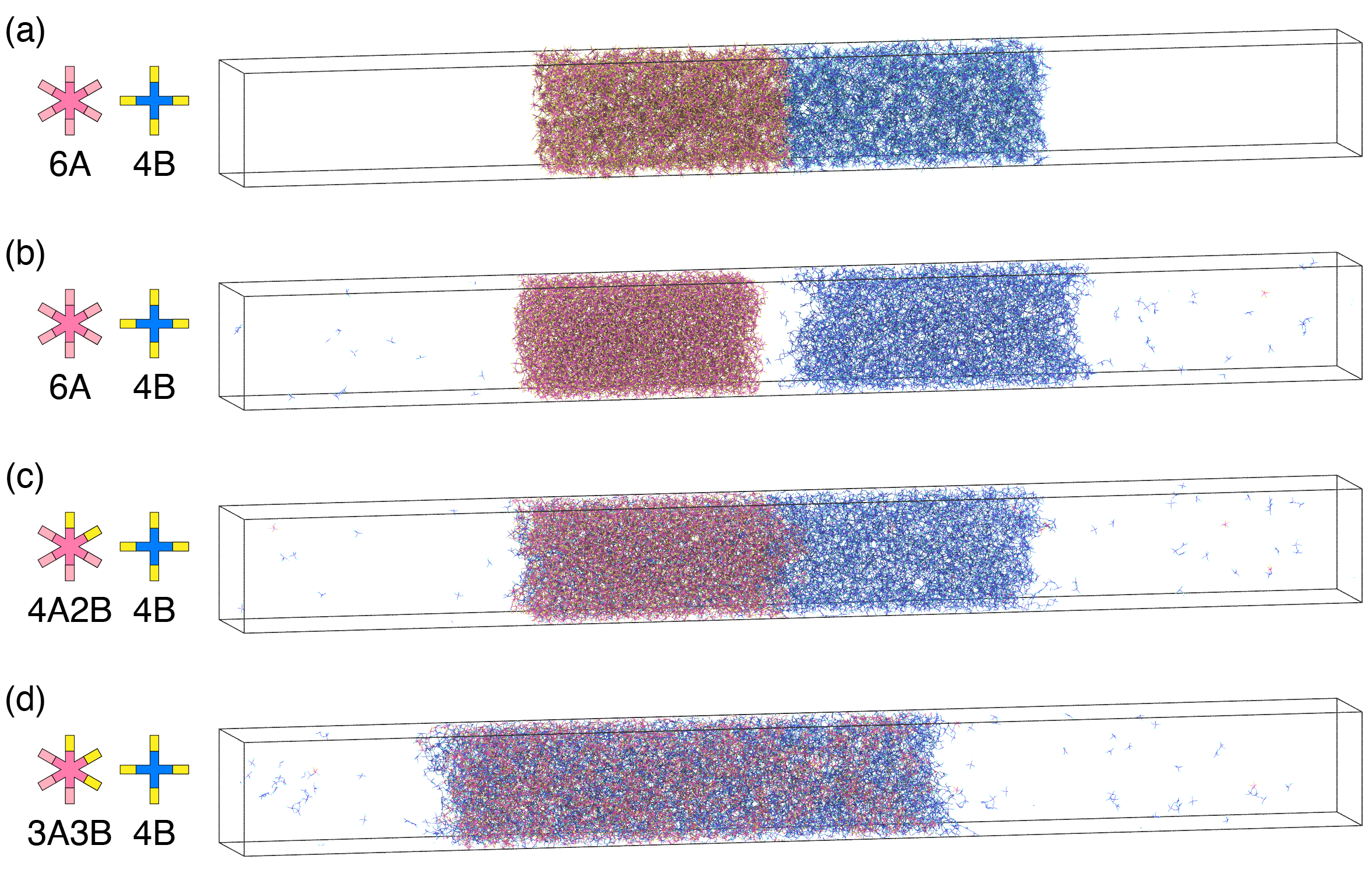}
	\caption{Simulation snapshots of two-component systems to calculate interfacial energy. The simulation box size is $1000\times100\times100$~nm$^3$ and the periodic boundary is applied in all directions. (a) Initial configuration for 6A + 4B and all other simulations, where two types of components are randomly distributed in two blocks.  (b) Combination of 6A + 4B. The two dense phases drift away and are separated by the dilute phase. (c) Combination of 4A2B + 4B. There is an interface between two dense phases. (d) Combination of 3A3B + 4B. There is only one mixed dense phase, which requires a very long time to equilibrate.}
	\label{figs2}
\end{figure*}

\renewcommand{\thefigure}{S\arabic{figure}}
\begin{figure*}[!htb]
	\centering
	\includegraphics[width=\linewidth]{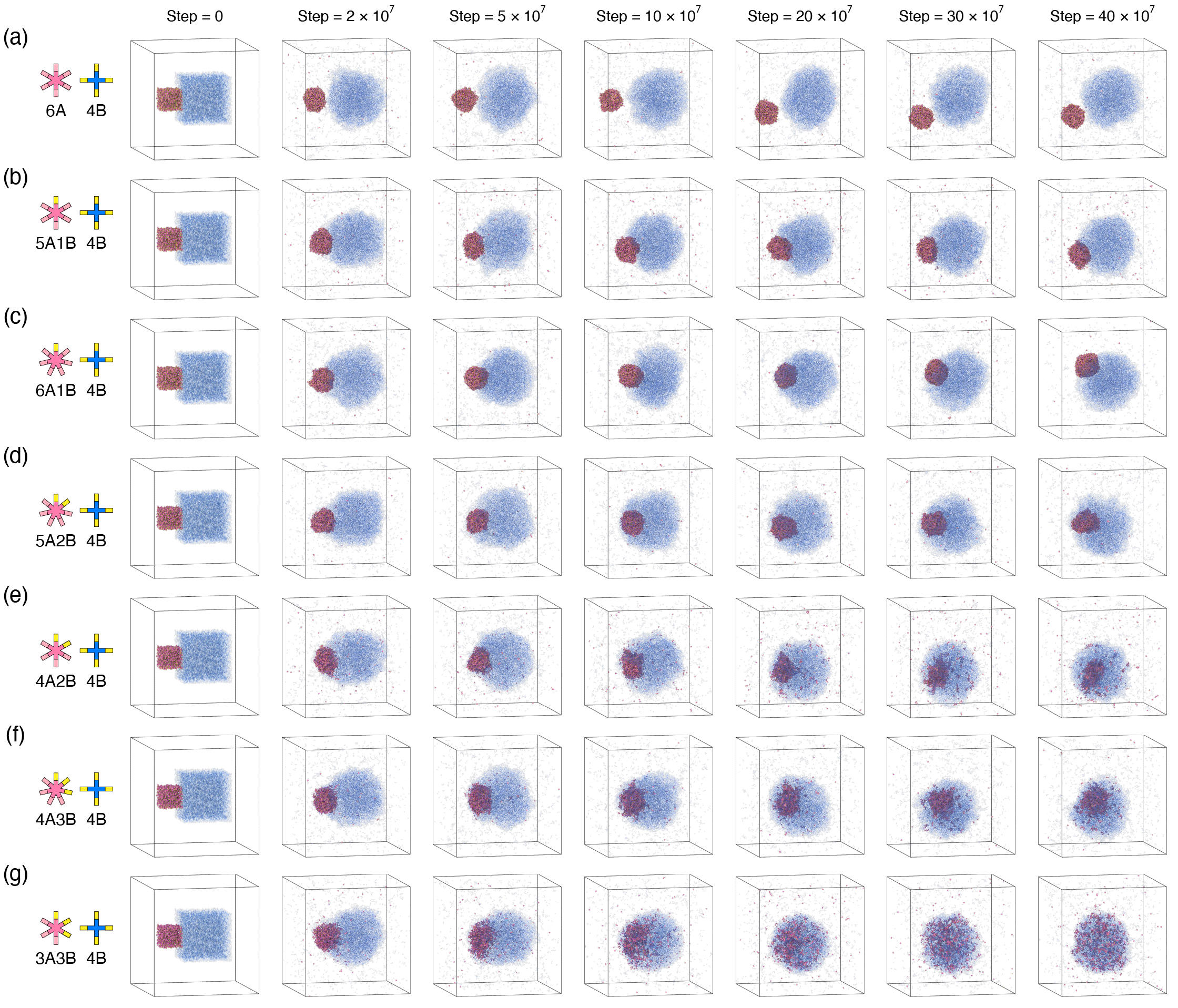}
	\caption{Designing different morphologies by using heterogeneous DNA nanostars. The simulation box size is $500\times500\times500$ nm$^3$ and the periodic boundary is applied in all directions. The numbers of Components 1 and 2 are 1500 and 20000, respectively. The type distributions of sticky ends are (a) 6A + 4B, (b) 5A1B + 4B, (c) 6A1B + 4B, (d) 5A2B + 4B, (e) 4A2B + 4B, (f) 4A3B + 4B, and (g) 3A3B + 4B, respectively. The binding energy of sticky ends is $\Delta E=-19k_{\rm B} T$.}
	\label{figs3}
\end{figure*}

\renewcommand{\thefigure}{S\arabic{figure}}
\begin{figure*}[!htb]
	\centering
	\includegraphics[width=\linewidth]{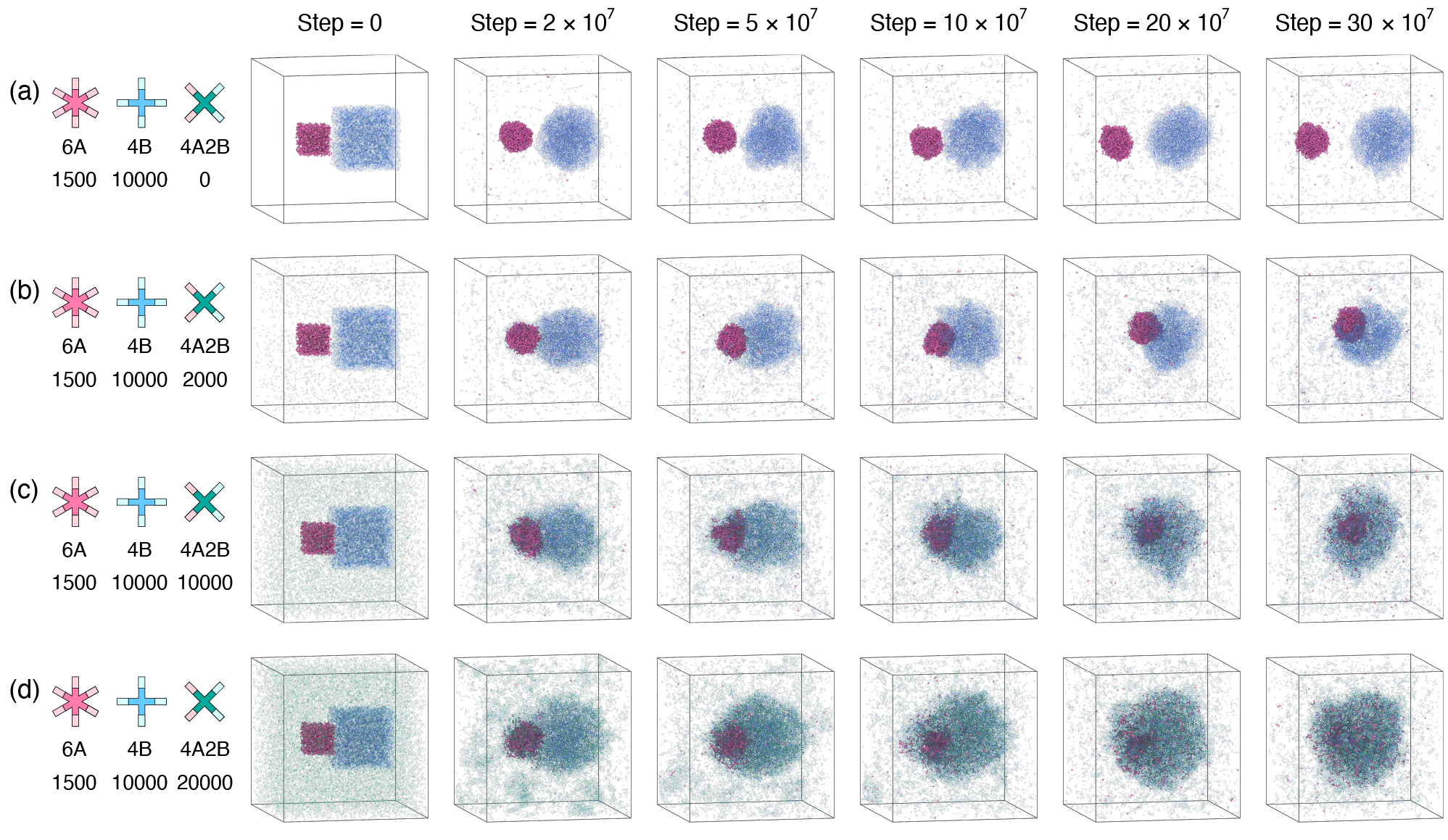}
	\caption{Designing different morphologies by introducing crosslinkers. The simulation box size is $500\times500\times500$ nm$^3$ and the periodic boundary is applied in all directions. The numbers of Components 1 and 2 are 1500 and 10000, respectively. The type distributions of sticky ends are 4A, 4B, and 2A2B for Component 1, 2, and the crosslinkers, respectively. The numbers of crosslinkers are (a) 0, (b) 2000, (c) 10000, and (d) 20000, respectively. The binding energy of sticky ends is $\Delta E=-19k_{\rm B} T$.}
	\label{figs4}
\end{figure*}

\renewcommand{\thefigure}{S\arabic{figure}}
\begin{figure*}[!htb]
	\centering
	\includegraphics[width=\linewidth]{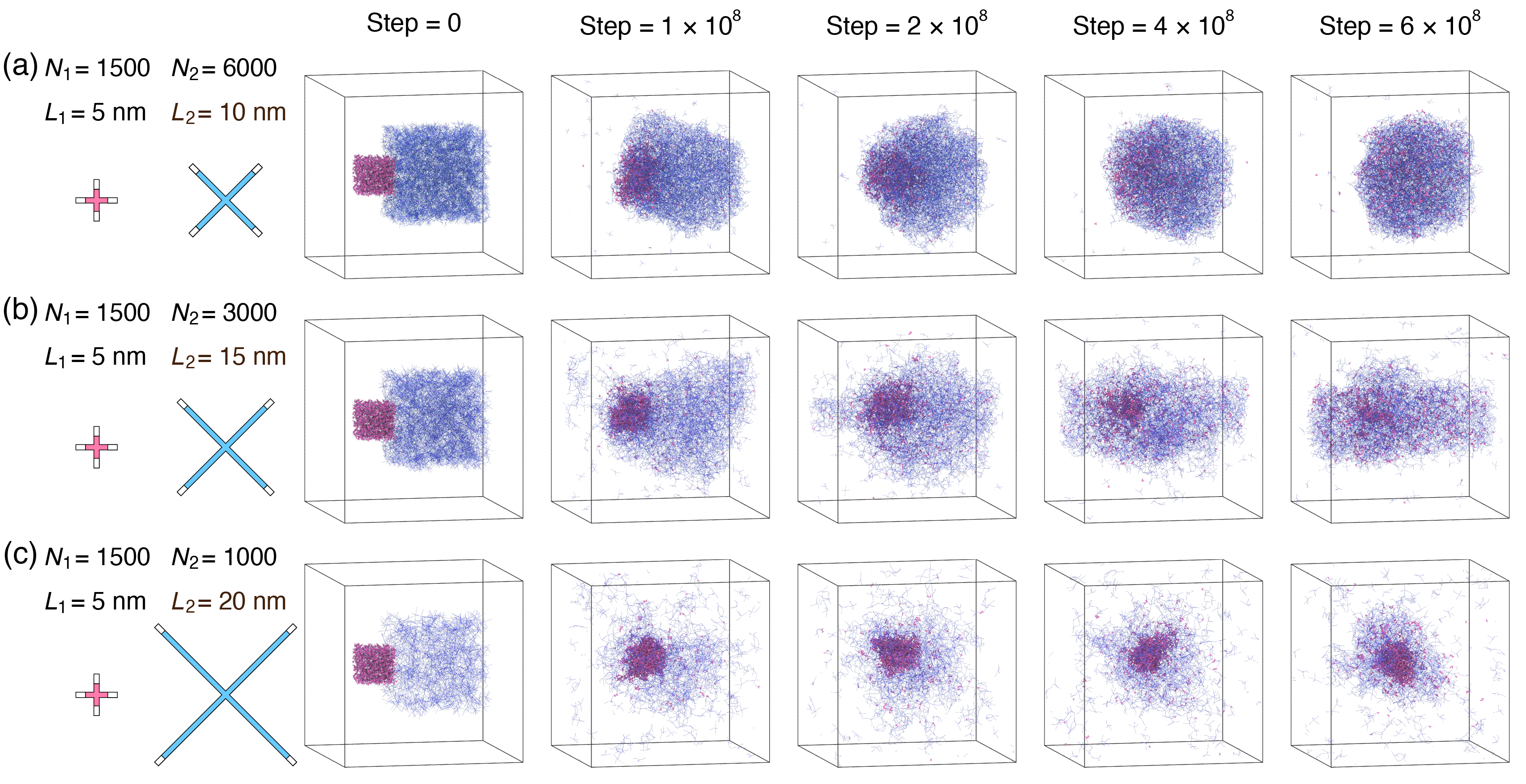}
	\caption{Designing different morphologies by using differently sized DNA nanostars. The simulation box size is $500\times500\times500$ nm$^3$ and the periodic boundary is applied in all directions. The type distributions of sticky ends are 4A + 4A, respectively. The arm length of Component 1 is 5 nm and the arm length of Component 2 is (a) 10 nm, (b) 15 nm, and (c) 20 nm, respectively. The number of Components 1 is 1500 for all cases. The number of Component 2 is (a) 6000, (b) 3000, and (c) 1000, respectively. The binding energy of sticky ends is $\Delta E=-22k_{\rm B} T$.}
	\label{figs5}
\end{figure*}

\renewcommand{\thefigure}{S\arabic{figure}}
\begin{figure*}[!htb]
	\centering
	\includegraphics[width=\linewidth]{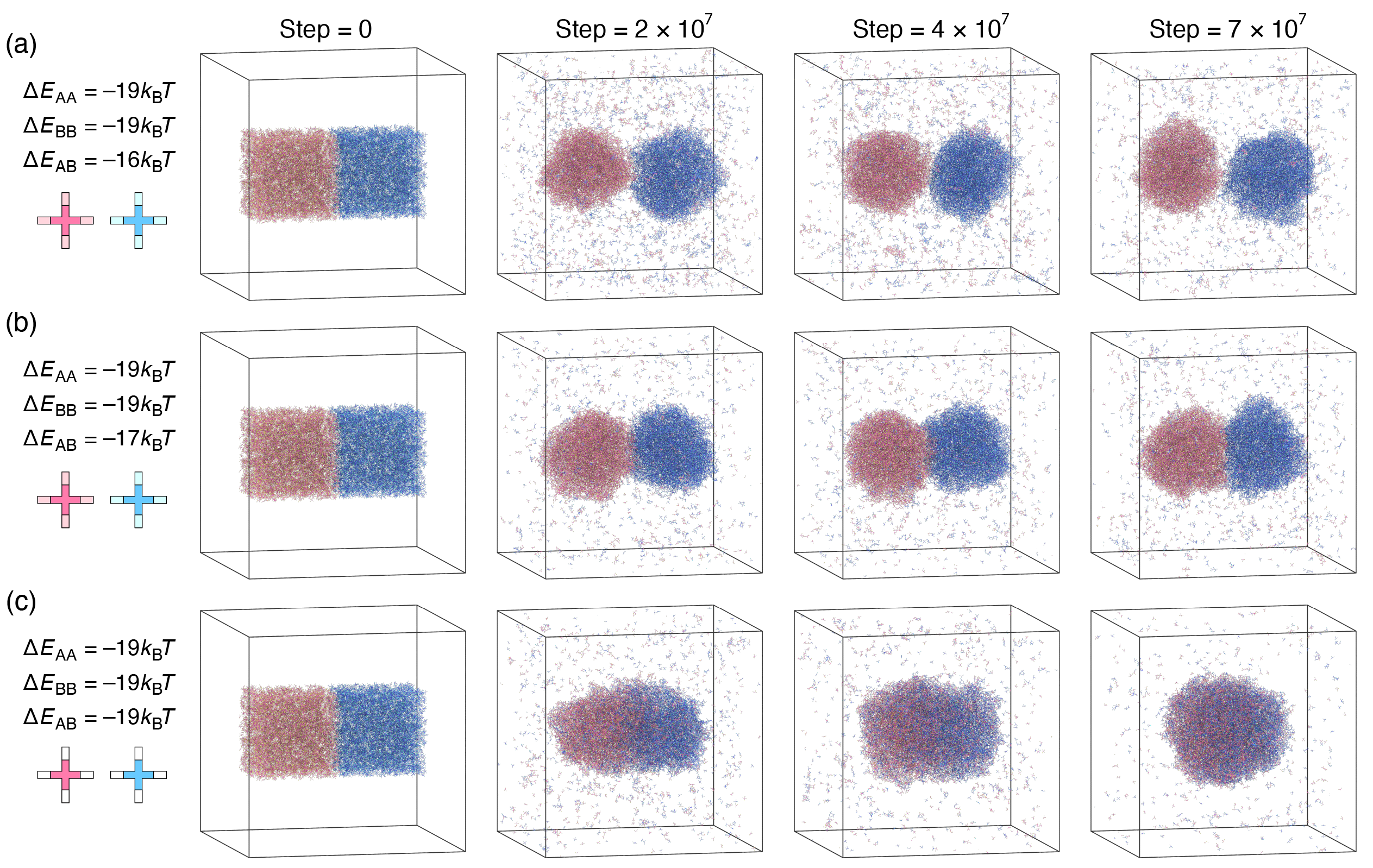}
	\caption{Designing different morphologies by changing the binding energy of sticky ends. The simulation box size is $500\times500\times500$ nm$^3$ and the periodic boundary is applied in all directions. The numbers of Components 1 and 2 are both 8000. The type distributions of sticky ends are 4A and 4B for Components 1 and 2, respectively. The binding energy of sticky ends is $\Delta E_{AA}= \Delta E_{BB}=-19k_{\rm B} T$. Here, we examine (a) $\Delta E_{AB}=-16k_{\rm B} T$, (b) $ \Delta E_{AB}=-17k_{\rm B} T$, and (c) $\Delta E_{AB}=-19k_{\rm B} T$ and find that the interaction between different types of sticky ends can also change the morphology of condensates. In all other parts of this work, we instead set $\Delta E_{AB}=100k_{\rm B}T$  to account for the steric repulsion between two non-identical sticky ends.}
	\label{figs6}
\end{figure*}

\renewcommand{\thetable}{S\arabic{table}}
\begin{table}[htbp]
    \centering
    \includegraphics[width=\linewidth]{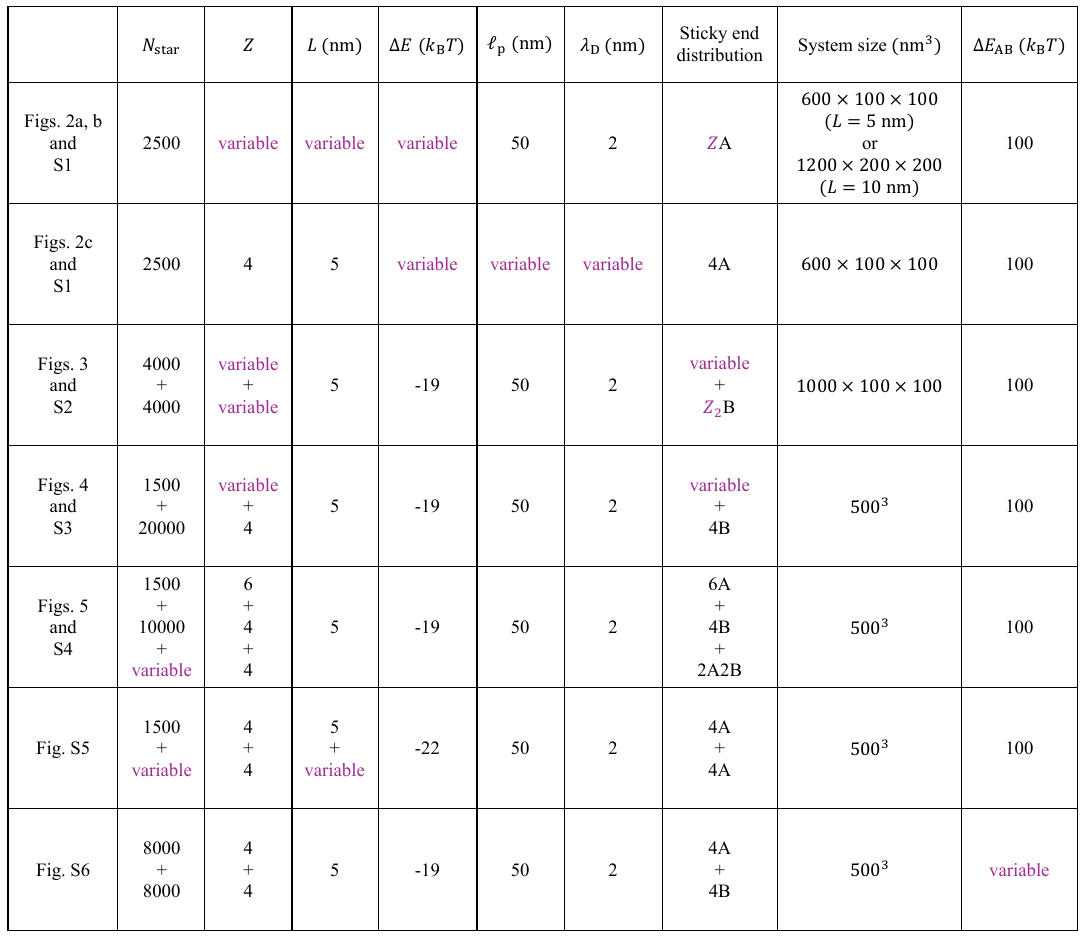}
    \caption{Summary of the molecular dynamics simulation parameters.}
    \label{table:md}
\end{table}